\documentclass[sigconf]{acmart}
\usepackage{graphicx}
\usepackage[T1]{fontenc}
\usepackage[utf8]{inputenc}
\usepackage{subcaption}

\usepackage{makecell}
\usepackage{multirow}
\usepackage{xcolor}
\usepackage{pifont}
\newcommand{\cmark}{\ding{51}} 

\AtBeginDocument{%
  }

\begin{document}

\title[Towards Considerate Embodied AI]{Towards Considerate Embodied AI: \\Co-Designing Situated Multi-Site Healthcare Robots from Abstract Concepts to High-Fidelity Prototypes}

\author{Yuanchen Bai}
\affiliation{%
  \institution{Cornell Tech / Cornell University}
  \city{New York / Ithaca}
  \state{New York}
  \country{USA}}
\email{yb299@cornell.edu}

\author{Ruixiang Han}
  \affiliation{%
  \institution{Cornell Tech / Cornell University}
  \city{New York / Ithaca}
  \state{New York}
  \country{USA}}
\email{rh652@cornell.edu}

\author{Niti Parikh}
\affiliation{%
  \institution{Cornell Tech}
  \city{New York}
  \state{New York}
  \country{USA}}
\email{ntp27@cornell.edu}

\author{Wendy Ju}
\affiliation{%
  \institution{Cornell Tech}
  \city{New York}
  \state{New York}
  \country{USA}}
\email{wendyju@cornell.edu}

\author{Angelique Taylor}
  \affiliation{%
  \institution{Cornell Tech / Cornell University}
  \city{New York / Ithaca}
  \state{New York}
  \country{USA}}
\email{amt298@cornell.edu}

\renewcommand{\shortauthors}{Bai et al.}

\begin{abstract}
Co-design is essential for grounding embodied artificial intelligence (AI) systems in real-world contexts, especially high-stakes domains such as healthcare. While prior work has explored multidisciplinary collaboration, iterative prototyping, and support for non-technical participants, few have interwoven these into a sustained co-design process. Such efforts often target one context and low-fidelity stages, limiting the generalizability of findings and obscuring how participants' ideas evolve. To address these limitations, we conducted a 14-week workshop with a multidisciplinary team of 22 participants, centered around how embodied AI can reduce non-value-added task burdens in three healthcare settings: emergency departments, rehabilitation facilities, and sleep disorder clinics. We found that the iterative progression from abstract brainstorming to high-fidelity prototypes, supported by educational scaffolds, enabled participants to understand real-world trade-offs and generate more deployable solutions. We propose eight guidelines for co-designing more considerate embodied AI: attuned to context, responsive to social dynamics, mindful of expectations, and grounded in deployment.
\end{abstract}

\keywords{Healthcare robots, Emergency department, Sleep disorder clinic, Long-term rehabilitation facility}

\begin{teaserfigure}
  \centering
  \includegraphics[width=0.9\textwidth]{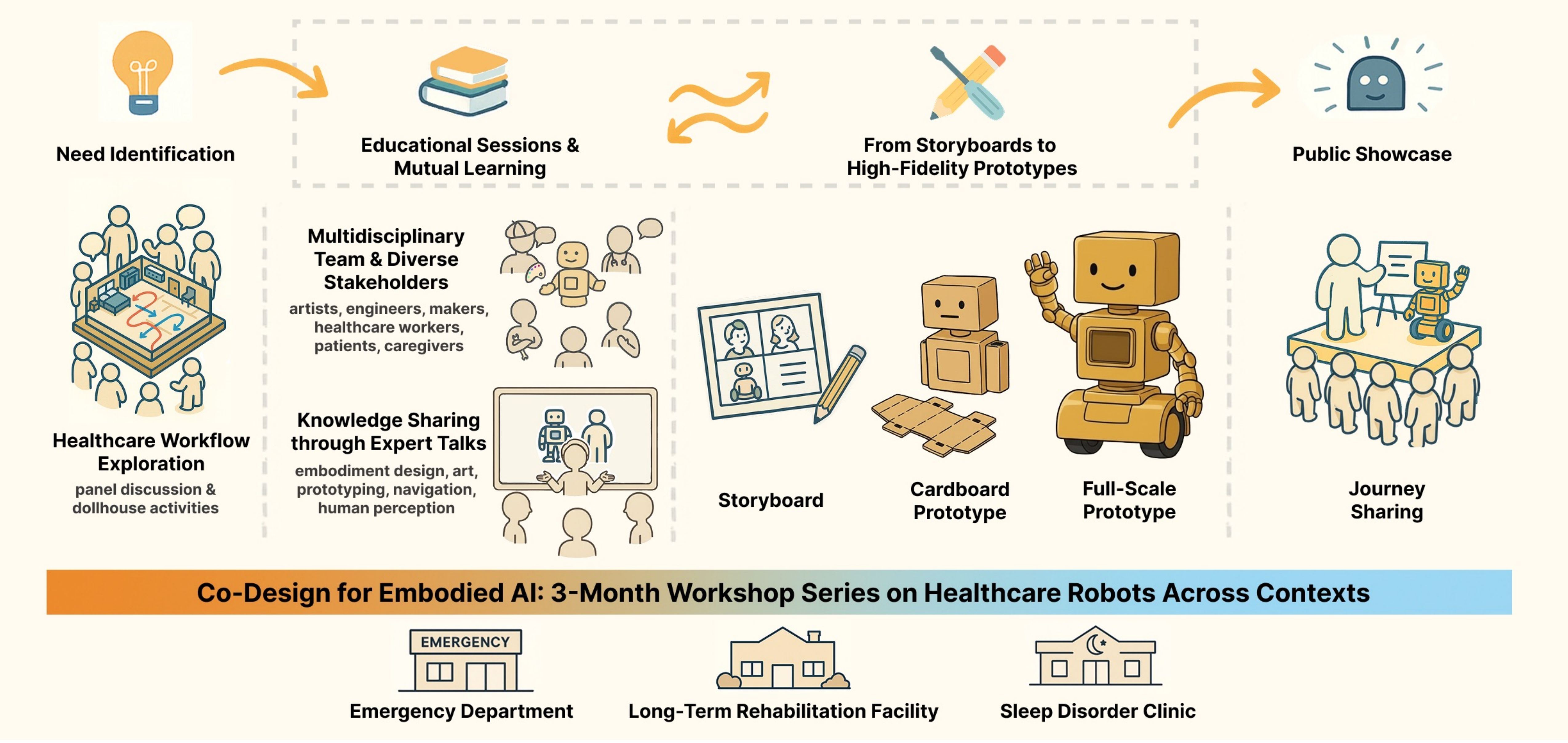}
  \caption{Overview of a three-month co-design workshop series for healthcare robots across three contexts: emergency department, long-term rehabilitation facility, and sleep disorder clinic. It progressed from need identification to prototyping and culminated in a final public showcase, with educational sessions scaffolding iterative refinement throughout.}
  \Description{Overview of a three-month co-design workshop series for healthcare robots across three contexts: emergency department, long-term rehabilitation facility and sleep disorder clinic. It progressed from need identification to prototyping and culminated in a final public showcase, with educational sessions scaffolding iterative refinement throughout.}
  \label{fig:teaser}
\end{teaserfigure}

\maketitle

\section{Introduction}

In recent years, artificial intelligence (AI) systems have seen growing exploration beyond disembodied assistants toward more physically grounded and interactive agents, commonly referred to as \textit{Embodied AI}, which are capable of perceiving their surroundings and engaging directly in real-world actions \cite{liu2025aligning}.
Embodied agents (e.g., robots) unlock greater potential for rich services and sensory experiences, while also introducing higher demands on perception, reasoning, planning, and action execution in response to the complexities of real-world environments and human needs \cite{liu2025aligning}.

Facing rising demands in healthcare, robots are increasingly envisioned to reduce healthcare workers' (HCWs) burden and improve patient care \cite{riek2017healthcare,hoedl2021influence}. 
Their roles span logistical delivery \cite{bloss2011mobile,pinna2015drug}, rehabilitation assistance \cite{topping2002overview,xie2021hybrid,dong2021state}, and emotional support \cite{abdi2018scoping,ikeuchi2018utilizing}. 
Non-value-added (NVA) tasks, such as excessive paperwork, avoidable movement, and redundant communication, can heighten stress for HCWs and distract from core responsibility \cite{capuano2004work,deans2011finding,nonvalueadded},  prompting growing interest in exploring robots to reduce operational inefficiencies and refocus attention on patient care.

However, embedding robots into healthcare goes far beyond technical automation.
Clinical environments are complex, with each facility imposing unique expectations for functionality, navigation, and interaction \cite{trainum2024nursing}: emergency departments (EDs) are fast-paced and chaotic \cite{taylor2020situating}, sleep disorder clinics (SDCs) demand personalized protocols for conducting studies \cite{sleepclinic}, and long-term rehabilitation (LTRs) centers must address intertwined medical and emotional needs \cite{sun2023data}. 
These variations highlight the diverse, context-sensitive forms of embodiment that robotic systems must take on. Moreover, the design of robots, including feasibility, cost, acceptability, and usability, all play a critical role in shaping their adoption \cite{silvera2024robotics,riek2017healthcare,taylor2019coordinating}. 
Addressing these factors requires close collaboration with frontline stakeholders and experts across disciplines.

Co-design enables such engagement by collaboratively designing with stakeholders, allowing researchers to access situated knowledge and user preferences that inform context-appropriate designs \cite{frijns2024co,rogers2022maximizing}. 
Prior HRI work has applied co-design in healthcare domains to promote physical activity \cite{antony2023co}, develop social robots for pediatric EDs \cite{foster2023co}, and support dementia care \cite{moharana2019robots}. 

However, several research gaps remain. 
First, many co-design efforts lack contextual depth and often focus on a single healthcare setting, limiting understanding of how robot roles might generalize, or need to specialize, across environments. 
They also tend to emphasize needs identification and preference elicitation, with relatively limited integration of multi-disciplinary input during solution generation. Second, current co-design efforts remain at lower-fidelity stages (e.g., storyboards), without progressing toward higher-fidelity prototypes that better approximate real-world deployment. 
This gap leaves the progression of participant understanding, from abstract ideas to embodied feasibility, largely unexplored. 
Third, prior work often engages stakeholders separately and defers synthesis to researchers post hoc, missing opportunities for collaborative exchange across roles and disciplines, such as healthcare, robotics, and design. 
Finally, there remains limited guidance on how to scaffold education for non-technical participants in understanding and contributing to robot design.

Addressing these gaps is crucial. Without cross-context learning, we lack the depth of contextual understanding needed to distinguish between generalizable patterns and context-specific requirements, hindering both theoretical advancement and practical deployment of context-aware robotic systems. 
Besides, high-fidelity prototyping is not merely about visual realism; it plays a critical role in surfacing real-world deployment tensions, such as spatial constraints, navigational conflicts, and human discomfort, that cannot be fully perceived or addressed through low-fidelity, conceptual design stages.
Moreover, healthcare robots' function, form, and interaction style must reflect interdisciplinary insight, from engineering and clinical practice to aesthetics and care ethics. 
In this light, co-design for embodied AI is not merely a method for building robots, but a way to define how they should be situated. Such robots must be \emph{considerate}: sensitive to context, responsive to social dynamics, aligned with situated goals, and mindful of stakeholder expectations.

Building on this, recent advances in multimodal reasoning \cite{wang2025multimodal}, world modeling \cite{zhou2024dino}, and edge computing \cite{tahir2025edge} have inspired researchers across domains (e.g., industry construction \cite{SUN2026106671}, surgery \cite{atar2025humanoidshospitalstechnicalstudy}, agriculture \cite{spagnuolo2025agricultural}) to explore how more capable embodied AI can support complex real-world tasks. This shift reflects a broader re-envisioning of embodied AI, toward more varied embodiment forms, more creative interaction activities, and more diverse use cases across everyday life and work settings. To fully tap these possibilities, we now need to consider a wider embodied-AI design space, including workflow fit as well as spatial, temporal, and proxemic constraints. Such an expanded perspective extends beyond what prior co-design guidelines (i.e., based on stationary, conversational social robots \cite{ostrowski2021long}) were able to address.

To fill these gaps, we organized a 14-week series of community-based co-design workshops, each lasting three hours per week (see Figure~\ref{fig:workshop_overview}).
Starting with needs identification and workflow mapping, the workshops advanced through iterative prototyping, spanning from storyboards to high-fidelity full-scale prototypes (see Figure~\ref{fig:artifacts}).
These activities were scaffolded with educational sessions designed to equip participants with diverse levels of robotics expertise to engage effectively in collaborative design.
By recruiting HCWs from the target facilities and involving interdisciplinary participants, we developed workflow-fit robotic solutions for three targeted facilities across distinct healthcare settings: ED, SDC, and LTR.

Our study was guided by the following four research questions:
\begin{itemize}
\item \textbf{RQ1:} What are the similarities and differences in robot roles to support NVA tasks across EDs, LTRs, and SDCs?
\item \textbf{RQ2:} How does sustained, close collaboration among participants from different disciplinary backgrounds and diverse stakeholder groups influence co-design processes and workshop outcomes?
\item \textbf{RQ3:} How do hands-on, iterative co-ideation activities (i.e., transitioning from sketches to cardboard prototypes and full-scale interactive prototypes) support design refinement and real-world applicability of the proposed robotic solutions?
\item \textbf{RQ4:} How can educational sessions be structured to support participants with varying levels of technical expertise, particularly non-technical stakeholders, in robot design?
\end{itemize}

We make the following five contributions:

\begin{itemize}
  \item \textbf{Contextual Task Analysis}: We identified and analyzed key NVA tasks in three healthcare contexts (ED, LTR, SDC), forming the foundation for context-adaptive robotic deployment \textit{(see Figure~\ref{fig:artifacts}}).

  \item \textbf{Cross-Disciplinary Co-Design}: We examined how multidisciplinary teams collaboratively shaped robotic design directions across varied settings, highlighting the importance of shared domain understanding \textit{(see Table \ref{tab:results})}.

  \item \textbf{Evolution of Design Ideas}: We analyzed how participants' ideas evolved through iterative prototyping (i.e., storyboards, cardboard, and full-scale prototypes), revealing patterns of creative negotiation and contextual anchoring \textit{(see Table \ref{tab:results})}.

  \item \textbf{Scaffolded Technical Engagement}: We analyzed how different educational formats (e.g., demonstrations, technology advancement histories, and implementation walkthroughs) supported participants in contributing to technical design, while revealing their differing preferences across formats \textit{(see Table \ref{tab:results})}.

\item \textbf{Design Guidelines for Considerate Embodied AI}: We proposed eight co-design guidelines to support future development of considerate embodied AI systems across four dimensions: Embodied Needs Grounding, Embodied Constraints \& Feasibility, Embodied Literacy Building, and Embodied Design Space Expansion \textit{(see Table \ref{tab:research_guidelines})}.
\end{itemize}

\section{Background}

\begin{table*}[t]
\centering
\caption{
Comparison of co-design workshops for healthcare and clinical-support robots studies across four methodological categories 
(participant/involvement, artifact fidelity, educational scaffolding, and guidelines), 
comprising nine analytical dimensions.
\cmark{} = yes (blank = no).
MCxt. = multi-context engagement enabling cross-context learning; 
MDisc. = multi-disciplinary participant groups; 
CSus. = sustained cross-role, cross-disciplinary close collaboration across needs identification, co-design, and evaluation stages; 
Iter. = iterated prototyping across workshops; 
Hi-Fi. = high-fidelity embodied artifacts generated; 
Scaf. = structured educational scaffolding; 
Resp. = responsive scaffolding across workshop stages; 
Learn. = learning-trajectory and learning-outcome analysis; 
Guide. = explicit co-design guidelines proposed.
}
\label{tab:codesign_comparison}
\begin{tabular}{p{3.35cm} p{1.6cm}ccccccccc}
\toprule
& 
& \multicolumn{3}{c}{\textbf{Participant/Involvement}} 
& \multicolumn{2}{c}{\textbf{Artifact Fidelity}} 
& \multicolumn{3}{c}{\textbf{Educational Scaffolding}} 
& \textbf{Guideline} \\
\cmidrule(lr){3-5} \cmidrule(lr){6-7} \cmidrule(lr){8-10} \cmidrule(lr){11-11} 

\textbf{Paper} 
& \textbf{Context}
& \textbf{MCxt.} 
& \textbf{MDisc.} 
& \textbf{CSus.} 
& \textbf{Iter.} 
& \textbf{Hi-Fi.} 
& \textbf{Scaf.} 
& \textbf{Resp.} 
& \textbf{Learn.} 
& \textbf{Guide.} \\
\midrule

Ostrowski et al. ~(2021) \cite{ostrowski2021long}
& Eldercare
&  &  &  &  &  &  &  &  & \cmark \\

Antony et al. (2023) \cite{antony2023co}
& Eldercare
&  & \cmark &  &  &  &  &  &  &  \\

Foster et al. (2023) \cite{foster2023co}
& Pediatric ED
&  & \cmark &  &  &  &  &  &  &  \\

Hsu et al. (2024) \cite{hsu2024give}
& Eldercare
&  &  &  & \cmark &  & \cmark & \cmark & \cmark & \cmark \\

Frijns et al. (2024) \cite{frijns2024co}
& Eldercare
&  & \cmark &  &  &  & \cmark &  &  &  \\

\midrule
\textbf{This Paper} (2026)
& ED/LTR/SDC
& \cmark & \cmark & \cmark 
& \cmark & \cmark 
& \cmark & \cmark & \cmark 
& \cmark \\
\bottomrule

\end{tabular}
\end{table*}
 
\subsection{Needs, Efforts, and Challenges of Healthcare Robots Assisting HCWs}

Healthcare systems face increasing pressure due to an aging population, rising disease burdens, and growing incidence of medical errors \cite{vos2015global,silvera2024robotics}.
The resulting gap between care demands and limited resources has intensified HCW workloads, harming their physical and mental well-being \cite{riek2017healthcare,townsley2023healthcare,hoedl2021influence,scheunemann2011ethics,mcginn2019meeting}, which can in turn affect care quality and timeliness \cite{vos2015global,silvera2024robotics}.
These issues are further complicated by the diversity of care settings, each shaped by differences in patient populations, clinical workflows, and built environments \cite{taylor2022hospitals}.
For example, in LTR, where residents are often older adults or have physical or cognitive impairments, care is complex and fragmented, requiring HCWs to provide both daily clinical care and emotional support \cite{sun2023data,ullal2024iterative,trainum2024nursing,doty1985overview}.
In ED, HCWs face intense physical and emotional pressure while managing chaotic environments, with stressors such as heavy workloads, high patient acuity, safety-critical cases, and external pressures like patient wait times and shifting tasks \cite{taylor2024towards,healy2011stress,crilly2017measuring,johnston2016staff}.
In SDC, overnight and daytime sleep studies are conducted where HCWs greet patients, set up monitoring equipment, and help them adapt to personalized treatments \cite{sleepclinic}.
While some tasks require direct attention from HCWs, other tasks do not require medical training, taking HCWs away from critical patient care \cite{silvera2024robotics}.

Robots have demonstrated potential to reduce HCW workload and improve patient outcomes.
Applications span logistics, daily care, rehabilitation, psychosocial support, and clinical procedures \cite{silvera2024robotics}.
For example, in inpatient and outpatient facilities, \textit{Gary} \cite{Gary1} assist with housekeeping chores, \textit{TUG} \cite{intelrealsense2024} transports items across departments, and \textit{Moxi}~\cite{moxi2024} supports nurses by delivering medical supplies. 
In long-term care environments, \textit{Lokomat} \cite{Lokomat1} provides physical assistance in rehabilitation training, \textit{PARO} \cite{PARO1} offers social companionship, and \textit{Lio} \cite{Lio1} engages patients through entertainment while also handling routine tasks.

Despite recent advancements, the design and adoption of healthcare robotics remain challenging as robots need to be tailored to specific care settings, tasks, and real-world environments \cite{riek2017healthcare,taylor2021social}. 
Our work contributes to a deep exploration of real-life HCW workflows in EDs, LTRs, and SDCs.  
We identify NVA tasks, examine their role in workflows, and explore robotic solutions that integrate seamlessly into real-world contexts, to maximize adoption potential and empower stakeholders.
Cross-setting analysis further enables transferable insights and supports context-aware design.

\subsection{Importance of Integrating Diverse Stakeholder Perspectives in Healthcare Robot Design}

Prior work emphasizes that HCWs' acceptance, trust, perceived usability, and operational practicality are critical for adoption \cite{silvera2024robotics,trainum2024nursing}. 
The design of robots should align with the real-life clinical and patient workflows; otherwise, robots can inadvertently increase workloads and cause frustration \cite{trainum2024nursing}.
Morever, in-person care involves more than task completion \cite{trainum2024nursing}.
For instance, assisting a patient with a shower is not merely functional; it enables communication, offers emotional support, and allows HCWs to monitor skin changes indicative of health status \cite{klein2018robotic}.
These nuanced forms of care are well understood by HCWs but can be overlooked if robot design is only centered around functionality.
While HCWs are often included in the design process, they are typically consulted as domain experts or asked to provide feedback on final designs \cite{frijns2024co,foster2023co,antony2023co,randall2018engaging}, with limited opportunities to engage in full-cycle, iterative co-design.

Besides HCWs, involving other community members is crucial for the successful deployment and broader acceptance of healthcare robots \cite{trainum2024nursing,riek2017healthcare,flandorfer2012population}.
These stakeholders bring diverse needs, skills, and perspectives.
For instance, HCWs prioritize patient safety, while patients may feel that monitoring technologies lead to feelings of being surveilled (i.e., panopticon \cite{kuiler2023panopticon,frijns2024co}). 
Similarly, medical students emphasize the importance of privacy and nursing students highlight the value of social interaction features \cite{trainum2024nursing}.
Participants also contribute by sharing insights grounded in lived experience, as individuals are experts in their own lives \cite{lu2023participatory}.
Moreover, increasing users' technological literacy and awareness of robotic potential can further enhance acceptance and meaningful engagement \cite{abanovi2015ARO}.

Our study extends prior work by centering diverse stakeholders, including healthcare stakeholders, engineers, and artists, not merely as consultants, but as active co-designers.
Through long-term, cross-disciplinary collaboration, we surfaced functional needs and relational aspects of care, informing design directions that are more grounded and acceptable in real-world care contexts.

\subsection{Bridging Perspectives: Workshops for Community-Driven Co-Ideations}

In this work, we use long-term co-design workshops to support a wide range of activities and draw insights from participants. 
We address \textbf{four research gaps} in organizing long-term workshops for healthcare robotics (see Table \ref{tab:codesign_comparison}).

\textbf{Context, Participant / Involvement: the need for long-term, close co-ideation across stakeholder groups with diverse roles and expertise. } 
Prior workshops often separate stakeholder groups.
For example, physical therapists and older adults participated in different sessions when co-designing for physical activity promotion \cite{antony2023co}.
In designing social robots for pediatric EDs, healthcare professionals and caregivers were also consulted separately \cite{foster2023co}.
Prior studies often involve stakeholder groups in separate sessions and rely on researchers to later synthesize their perspectives into design insights \cite{antony2023co,foster2023co}, rather than fostering direct collaboration across groups from the outset.
While a recent study introduced a ``mixed-group'' format consisting of five workshops with residents and care workers to explore robotic technology in care \cite{frijns2024co}, our work further extends this direction by involving a broader range of stakeholder roles and expertises, and sustaining long-term, tightly integrated co-ideation, with researchers facilitating close communication and mutual learning across sessions. Moreover, cross-context comparison revealed both shared and context-specific needs, enabling a more nuanced interpretation of robot roles and the workflow implications of different design choices.

\textbf{Artifact Fidelity: tracing design progression through multi-stage artifact development. }
HRI researchers have used diverse methods (e.g. focus groups \cite{kitzinger1994methodology}, storyboarding \cite{truong2006storyboarding}, collaborative mapping \cite{lee2017collaborative}, prototyping \cite{steen2013co}, and panels \cite{hsu2024give}) to enable learning through design \cite{hsu2024give}.
Rather than relying on a single method or stopping at low-fidelity sketches, we scaffolded co-design through multiple stages (i.e. storyboards, cardboard prototypes, and full-scale prototypes), and analyzed how participants' ideas evolve through group discussions, interviews, and artifact analyses.

\textbf{Educational Scaffolding: the need for educational scaffolding to support informed co-design engagement. } 
Supporting meaningful participation in robot co-design requires informing participants about technical capabilities.
However, \cite{frijns2024co} points out that the current common practice of introducing predefined robot embodiments and application scenarios may limit participants' imagination about robot roles, and instead explores introducing technical modular components like conversational AI and computer vision to broaden creative thinking.
Our work goes one step further by structuring dedicated educational sessions across multiple domains (e.g., aesthetics, prototyping, robot technical capabilities), and analyzing how such sessions shape participants' expectations, co-ideation capacity, and perceptions of trade-offs between concrete demonstrations and open-ended exploration.

\textbf{Guideline: proposing long-term co-design guidelines to the healthcare HRI context.}
Existing long-term co-design guidelines (e.g., with older adults [44]) largely focus on interaction-level refinement with existing social robots, such as conversational or behavior-based iterations on low-degree-of-freedom (low-DoF) platforms. Our work extends beyond these scopes by situating long-term co-design in low- and high-stakes healthcare settings, and engaging multiple contexts (SDC, ED, LTR) with artifacts that evolve from low- to high-fidelity. Building on more advanced embodied-AI capabilities, our multi-stage process enabled richer envisioning of functions, activities, and embodiments, as well as long-term tracing of participants' learning and idea evolution. Together, these insights allow us to propose transferable, multi-dimensional guidelines that integrate learning, co-ideation, artifact refinement, and workshop organization for embodied-AI co-design.

\begin{figure*}[!htbp]
    \centering
    \includegraphics[width=1.0\textwidth]{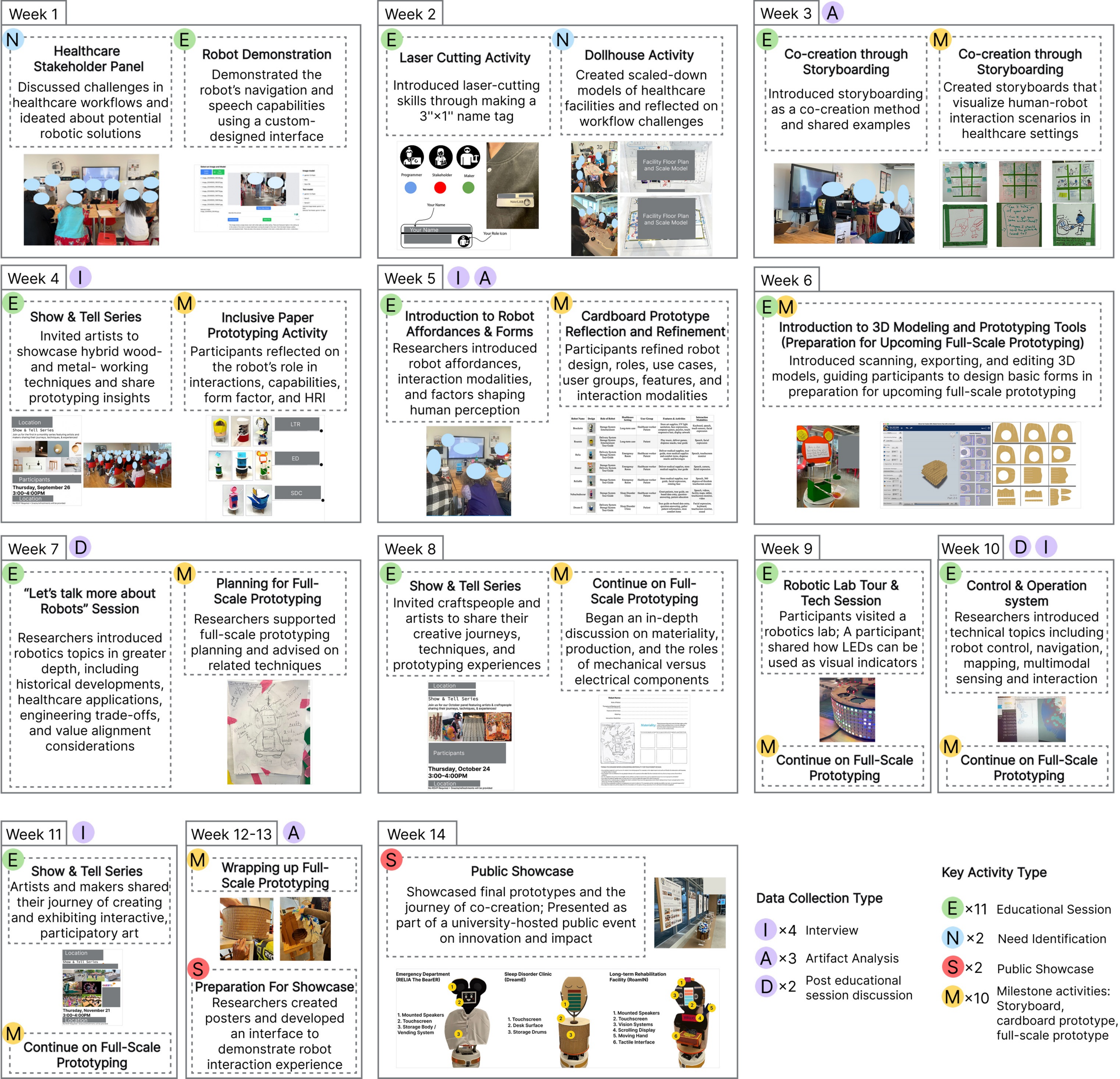}
    \caption{Overview of the workshop. Weekly activities spanned diverse formats, including educational sessions (E), need identification (N), milestone activities (M) (e.g., storyboard, cardboard prototype, full-scale prototype), and a public showcase (S). Colored tags on the top-left corners of each activity card indicate the key activity types. Additionally, we annotated each week with relevant data collection types, including interviews (I), artifact analysis (A), and post-educational-session discussions (D), to highlight the integration of research through co-creation.}
    \Description{A visual timeline of a 14-week co-design workshop illustrating the progression of activities from stakeholder discussions and early ideation to technical prototyping and a public showcase. Each week includes icon-labeled modules representing educational content, design milestones, and collaborative sessions. Tags in the top-left corners denote key activity types (e.g., E: educational, N: need identification, M: milestone, S: showcase). Weekly data collection methods are marked with purple symbols (e.g., I, A, D) to show where interviews, artifact analysis, and session discussions took place.}
    \label{fig:workshop_overview}
\end{figure*}

\begin{table*}[t]
\caption{Workshop participants, their self-identified roles, self-reported healthcare worker (HCW) experience where applicable, and their assigned facility-context co-design teams (ED, LTR, SDC). HCW experience is reported exactly as self-identified by participants. An asterisk (*) indicates an HCW who works directly within the facility for which their team was designing.}
\label{tab:participants}
\centering
\begin{tabular}{p{0.6cm} p{1cm} p{1.7cm} p{4cm} p{5cm} p{1.6cm}}
\toprule
\textbf{ID} & \textbf{Gender} & \textbf{Age} & \textbf{Role} &
\textbf{HCW Expertise \& Experience} & \textbf{Team} \\
\midrule
P1  & Female & 30--40 & \textbf{HCW} / Programmer & Microbiology \& Immunology; 0--5 yrs & ED \\
P2  & Female & 50--60 & \textbf{HCW*} & Chief of Sleep Medicine; $>$20 yrs & SDC \\
P3  & Male   & 60--70 & Programmer & N/A & ED \\
P4  & Female & 20--30 & \textbf{HCW*} / Programmer & Medical student with Sleep Medicine research experience; 5--10 yrs & SDC \\
P5  & Female & $>$70  & Artist & N/A & LTR \\
P6  & Female & 50--60 & \textbf{HCW} / Artist & Long-term Caregiver; 15--20 yrs & LTR \\
P7  & Female & 20--30 & Programmer & N/A & ED \\
P8  & Male   & 50--60 & Architect / Artist / LTR Patient & N/A & LTR, SDC \\
P9  & Male   & 60--70 & Artist & N/A & LTR, SDC \\
P10 & Female & 30--40 & Maker & N/A & ED \\
P11 & Male   & 60--70 & Maker / LTR Patient & N/A & LTR \\
P12 & Female & 60--70 & Artist & N/A & LTR \\
P13 & Female & 50--60 & \textbf{HCW*} & LTR Art Therapist; $>$20 yrs & LTR \\
P14 & Male   & 20--30 & Programmer & N/A & ED \\
P15 & Male   & 20--30 & Programmer / Maker & N/A & SDC \\
P16 & Female & 20--30 & Programmer & N/A & ED \\
P17 & Male   & 60--70 & \textbf{HCW*} & Internal Medicine Physician; $>$20 yrs & ED \\
P18 & Male   & 30--40 & Programmer / Maker & N/A & LTR \\
P19 & Female & 40--50 & Maker & N/A & LTR \\
P20 & Male   & 50--60 & \textbf{HCW} & Not Specified; $>$20 yrs & LTR \\
P21 & Male   & 20--30 & Programmer / Maker & N/A & ED \\
P22 & Male   & Not disclosed & \textbf{HCW*} & Clinical Emergency Medicine; 10--15 yrs & ED \\
\bottomrule
\end{tabular}
\end{table*}

\section{Methodology}
We conducted an IRB-approved (IRB \#: IRB0145631), 14-week workshop series inspired by a community co-making collective \cite{zhao2024craft} in a MakerLab space on a university campus located in a large metropolitan area in the global north. The workshop offered a total of 42 hours of in-person, hands-on co-ideation (3 hours per week).
Our aim was to prototype healthcare robots for ED, LTR, and SDC, providing participants with the opportunity to design robots for environments they found most interesting.
We organized iterative design activities (e.g., storyboarding, prototyping), supported by scaffolded educational sessions, flexible attendance accommodations, and structured group sharing, to foster informed ideation, cross-disciplinary engagement, mutual learning, and a co-creation environment characterized by collaboration and inclusion.
Two to three members of our team, from a university MakerLab and robotics lab, organized and facilitated all workshop sessions.

\subsection{Workshop Participants Recruitment \& Demographics}
\label{sec:participants}

We recruited participants through word-of-mouth, university listserv advertising, and our professional networks. Before the workshops, we specifically recruited HCWs who work directly within the facilities for which the robots were being designed (ED, SDC, LTR), ensuring that the co-design process was grounded in their contextual experience and workflow knowledge. Because the study required long-term commitment, we communicated goals and expectations upfront to ensure that HCWs could commit throughout the workshop series.
Other participants were recruited on an ongoing basis during the workshop series. We welcomed community members who could contribute diverse perspectives, including HCWs, patients, artists, makers, and programmers. 
Demographic information was collected at onboarding, and a total of 22 participants completed the demographic intake (Table \ref{tab:participants}).
They included 11 females and 11 males, aged 20-70+.
5 participants were artists, 6 were makers, 9 were programmers, 8 were HCWs.
Regarding the targeted facilities, two HCWs were doctors working in the ED. Two HCWs worked in the SDC (the Chief of Sleep Medicine and a medical student). One HCW in the LTR was an art therapist, and two other participants were residents/patients at the LTR facility. All participants volunteered to take part in the study and received no compensation.

\begin{figure*}[!htbp]
    \centering
    \includegraphics[width=0.85\textwidth]{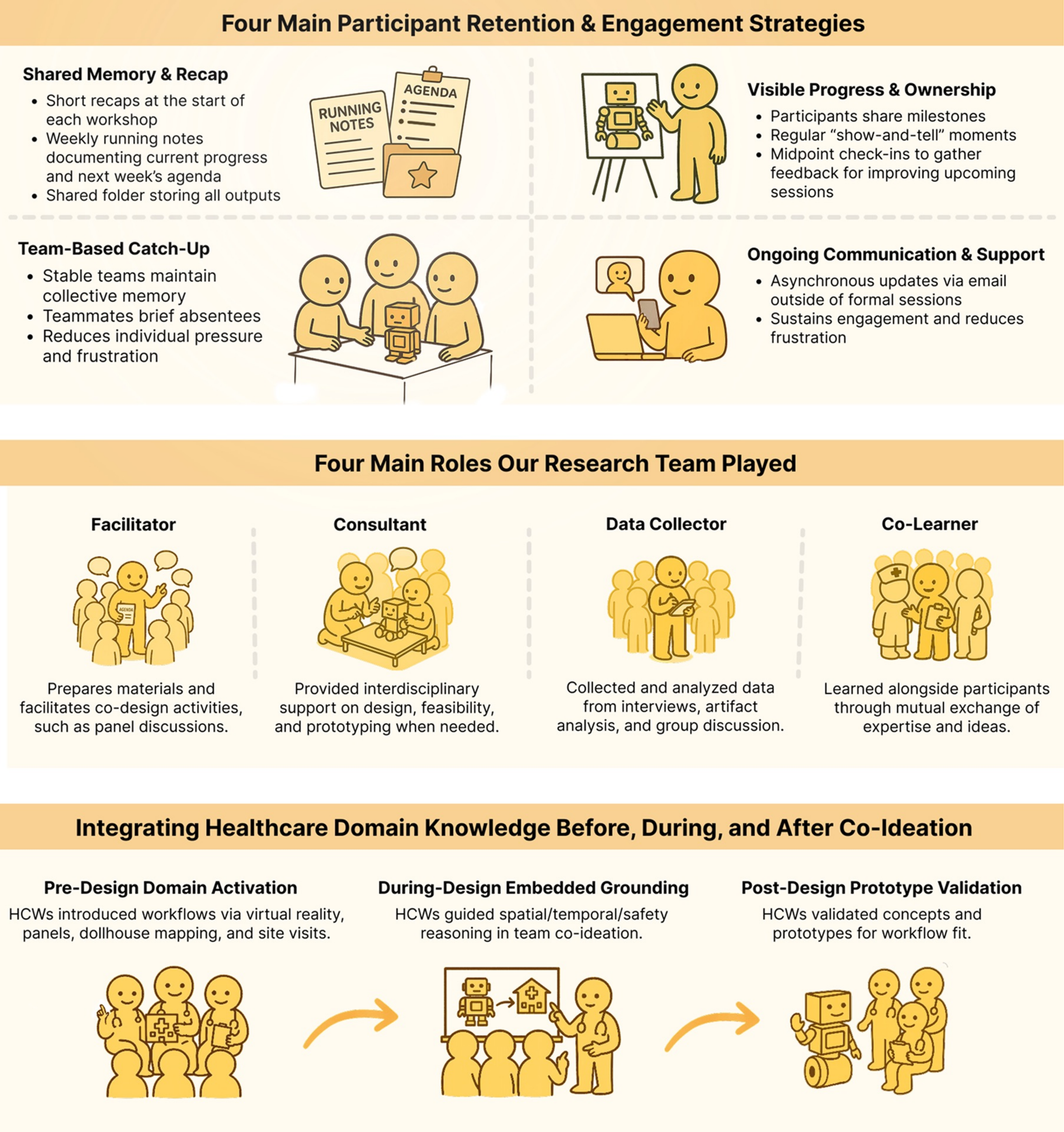}
    \caption{An overview of how long-term engagement, researcher involvement, and healthcare expertise were operationalized during the study: (1) four participant-retention strategies that supported continuity, re-engagement, and sustained momentum across sessions; (2) four roles enacted by the research team (i.e., facilitator, consultant, data collector, and co-learner) that structured involvement and maintained appropriate boundaries; and (3) the pre-design, during-design, and post-design integration of healthcare domain knowledge that guided design reasoning and informed prototype refinement.}
    \Description{This figure provides an overview of the three components that supported our long-term, high-fidelity co-design process: (1) participant retention and engagement strategies, (2) researcher roles and boundaries, and (3) healthcare domain knowledge integration across pre-design, during-design, and post-design phases, with iterative validation as prototypes evolved.}
    \label{fig:workshop_protocols}
\end{figure*}

\subsection{Workshop Platform and Equipment}
\label{sec:robot}

\textbf{Robotic Platform:} The workshop series was centered around augmenting a TIAGo Mobile Base robot with new embodiments for healthcare settings \cite{Tiago}.
TIAGo is well-suited for the workshop activities because it is an autonomous mobile platform with a payload capacity of 154.3 lb, a maximum speed of 2.24 mph, and a footprint measuring 21.3 inches in diameter by 11.8 inches in height, sufficient to carry new robot-body prototypes.
This robot provides a foundation for building various embodiments and attachments (e.g., 4 USB 2.0 ports, 2 USB 3.0 ports, Ethernet ports, GPIO pins, speaker), computing power (Intel Core i5 CPU, 256 GB SSD), and sensing capabilities (e.g., laser scanner) to perform a wide range of robotic tasks. 
We added additional compute and sensing functionality to the robot using a Jetson XAVIER GPU \cite{NVIDIA} and an OAK-D camera \cite{Camera} to demonstrate the robot's potential for situational awareness and responsiveness through diverse algorithms (e.g., computer vision). To support embodied co-design, we selected TIAGo as a mobile base that could anchor participants' imagination of a mobile robot with diverse embodiments and interaction capabilities. Participants could pick and place new robot forms onto the mobile base based on healthcare contexts, using it as a starting point for co-ideation.

\textbf{MakerLab Equipment:} The MakerLab is an open shared space for easy access to tools for designing, building, and testing new ideas from initial sketch to refined prototype.
The equipment used in the workshop supported participants in rapidly building and refining physical prototypes as their ideas progressed from sketches to full-scale designs, including 3D printers, CO2 laser cutting and engraving machines, soldering irons, handheld 3D scanners, and various woodworking tools.

\subsection{Standard Workshop Protocols}
\label{sec:protocal}

\subsubsection{Participant Retention and Attendance Documentation}

Given the long-term nature of the study and the potentially demanding schedules of participants, we expected potential occasional absences during preparation. Thus, to ensure that participants who missed a session could reenter the design process without losing context, we implemented four retention mechanisms (see Figure~\ref{fig:workshop_protocols}). We opened each workshop with a brief recap of prior sessions. Stable team configurations further supported recaps, as teammates could help one another catch up and sustain the momentum of ongoing ideas. To reinforce collective ownership, we kept evolving artifacts visible throughout the workshops, which helped participants reconnect with earlier contributions and track the group's progress. In addition, asynchronous communication between sessions provided timely updates and reduced frustration related to scheduling constraints. 
Regarding attendance documentation, the design work in this study was inherently group-based: ideas were developed, refined, and negotiated collectively within each team.  For each session, we recorded team artifacts, photographs, and videos of prototypes and discussions, enabling us to trace the evolution of each group's design ideas across the 14-week process. These group-level records, supported by our retention mechanisms, captured the collective evolution of ideas and also offered complementary information when individual attendance fluctuated.

\subsubsection{Researcher Roles and Influence Clarification}

Our research team took on four roles to support the workshops while maintaining transparency regarding their level of involvement and potential influence (see Figure~\ref{fig:workshop_protocols}). At different moments, our research team facilitated group discussions, offered interdisciplinary clarification when requested, documented interactions and artifacts, and learned alongside participants during collaborative reasoning tasks. These role boundaries supported participants in leading the generation and refinement of design ideas, while researchers provided structure only where necessary to sustain productive engagement and clarify well-established robot capabilities (e.g., navigation, verbal and non-verbal interaction in a given context).

\subsubsection{Participant Roles, Expertise, and Team Assignment}

We assigned participants to ED, LTR, or SDC co-design teams based on their interests and expertise.

\paragraph{1) Targeted HCWs and Domain-Grounded Co-Design}

HCWs recruited from the targeted facilities were placed in the team aligned with their clinical background (e.g., ED clinicians in the ED team; Sleep Medicine clinicians and researchers in the SDC team; the LTR art therapist in the LTR team). We organized activities so that their domain grounding shaped all stages of the design process (see Figure~\ref{fig:workshop_protocols}): pre-design activation (i.e, panel discussions, dollhouse activity mapping), during-design grounding (i.e., storyboarding, cardboard prototyping, full-scale robot prototyping), and post-design validation (i.e., iteratively reviewing evolving prototypes, including feature tables and summaries). 

\paragraph{2) Other HCWs \& Non-HCW Multidisciplinary Contributions}
Beyond the targeted HCWs, additional HCWs joined teams based on their interests, providing additional clinical standpoints.
Non-HCW participants, including artists, makers, programmers, and rehabilitation residents, brought creative, technical, and experiential perspectives that complemented HCW reasoning and needs assessments. While they primarily remained within a single facility team to support continuity, their specialized skills were shared across teams when relevant. For example, P9, a professional illustrator and theatre designer, produced storyboards in the LTR team and later supported the SDC team with lighting and ambience concepts for sleep-supportive environments. P8, a LTR resident with architectural training, contributed lived-experience insights to the LTR team and assisted with SDC prototyping when extra support was needed. These multidisciplinary contributions expanded the design space and enriched cross-team collaboration.

\subsection{Workshop Activities}
\label{sec:activities}
To support iterative co-design and scaffold participation across diverse stakeholder groups, we structured the 14-week community-based workshop series around four key activity types (see Figure \ref{fig:workshop_overview}): \textbf{Need Identification}, \textbf{Educational Sessions}, \textbf{Milestone Activities}, and a final \textbf{Public Showcase}. 
These activities were strategically distributed and interwoven to foster mutual learning, sustained engagement, and meaningful contribution.

\textbf{1)} In the early phases, we conducted two \textbf{Need Identification} sessions to surface challenges across stakeholder groups. 
The first (Week 1) was an HCW stakeholder panel, where participants posed questions and shared insights with representatives from the three stakeholder facilities. 
The second (Week 2) was a dollhouse activity, during which participants collaboratively explored 3D-printed miniature care environments and used scaled artifacts to discuss spatial and workflow challenges in daily operations.
These sessions helped ground the design in real-world contexts.

\textbf{2)} We also integrated three \textbf{Milestone Activities} to anchor collaborative ideation and design reflection. 
These included storyboarding, cardboard prototyping, and high-fidelity full-scale prototyping, all of which invited participants to externalize and iteratively refine their ideas. 
These activities served as shared reference points to maintain engagement and visualize progress.

\textbf{3)} To support participants from varied technical and professional backgrounds, we designed a series of \textbf{Educational Sessions} that addressed key areas such as robotics, design, fabrication, and art. 
These sessions were intended not only to introduce robotics but also to encourage participants to share their own domain knowledge. 
For example, our ``Show \& Tell'' series featured invited artists and makers who shared their creative processes, prototyping practices, and fabrication methods.
Researchers also led sessions on storyboarding and design techniques to support milestone activities.

Given the robotics-centered nature of the project, we placed particular emphasis on \textbf{robotics education}, progressively introducing key concepts to align with participants' evolving engagement levels. 
In Week 1, a hands-on demonstration of the TIAGo robot showcased integrated navigation and speech capabilities.
In Week 5, we introduced the concept of robot affordances and physical forms to support cardboard prototyping. 
Later sessions deepened the discussion. For example, the \textit{``Let's Talk More About Robotics''} session covered topics such as development history, programming, value alignment, and healthcare applications. 
In Weeks 9 and 10, participants explored robot navigation and environmental mapping.
These sessions were structured to address knowledge gaps and progressively deepen participants' understanding.

\textbf{4)} The final \textbf{Public Showcase} invited participants to present their full-scale prototypes and reflect on the co-creation journey. 
This event was part of a broader university-level initiative aimed at fostering innovation and community engagement across student and entrepreneurial communities, further amplifying the visibility and impact of the work.
\subsection{Data Collection}
\label{sec:datacollection}
As our workshop series includes tightly packed sessions and a wide range of evolving activities, we adopted multiple forms of data collection. 
We recorded audio and video throughout the sessions and documented participant engagement through various structured materials, such as worksheets, activity tables, collaborative notes, and emails. 
Our analysis draws from three primary sources: 

\textbf{1) Artifact Analysis.} We conducted analysis of three types of artifacts (i.e., storyboarding, cardboard prototyping, and full-scale prototyping) to explore how participants' ideas evolved over time and how their design concepts became progressively refined. 

\textbf{2) Interview.}  We conducted a total of four semi-structured interviews across different phases of the workshop, which can be grouped into two types. 
The first two participants were formal one-on-one interviews, scheduled according to
participants' availability and conducted in quiet settings during Weeks 4 and 5.
To ensure a positive experience and avoid disrupting prototyping work,
participation was kept entirely voluntary. For instance, P20
declined but recommended P19, whom they felt could offer more relevant insights. As a result, 11 participants were interviewed in this phase. This interview sample included participants from all three facility teams (ED, LTR, SDC) and covered HCWs, patients/residents, and artists/makers/engineers.
These interviews focused on participants' reflections on the stakeholder panel session, their experiences with multidisciplinary collaboration, and their impressions of the overall workshop organization.
Due to limited time and participants' intensive involvement in hands-on making in later stages, we were unable to schedule additional one-on-one interviews. 
Therefore, the remaining two were informal, focus-group-style interviews conducted during the prototyping sessions in Weeks 10 and 11, where we visited each of the three groups (i.e., ED, LTR, SDC) in their working spaces and engaged them in small-group discussions while they were prototyping.

\textbf{3) Post-Educational Session Discussion.} We analyzed discussions conducted immediately following two key educational sessions: (1) \textit{``Let's Talk More About Robotics'' } (Week 7) and (2) \textit{``Control and Operation System''} technical session (Week 9). 
These sessions provided space for reflection, clarification, and exchange of perspectives after participants had gained foundational knowledge. 

\subsection{Data Analysis.}
\textbf{1) RQ1}. We analyzed the three stages of design artifacts: storyboards, cardboard prototypes, and final full-scale prototypes. 
Storyboards were documented based on annotations from participants and facilitators for each group.
For cardboard prototypes, we prepared a table including attributes such as robot roles, interaction modalities, activities, and features for participants to complete (inspired by prior robot co-design work \cite{moharana2019robots}), and discussed these elements with group members for refinement.
Full-scale prototypes were summarized on posters annotated with details from participants' written documentation (e.g., emails) and in-session narratives.
All outcomes were reviewed and confirmed by participants to ensure accuracy.
For cross-context comparisons of robotic solutions, the storyboards captured early divergent brainstorming and were relatively abstract, whereas the cardboard prototypes followed one iteration and incorporated detailed discussions of modalities, features, and user considerations. 
In contrast, the full-scale prototypes reflected a prioritized integration of ideas, yet each team could realize only one consolidated design due to time constraints. 
Therefore, our main analysis focused on the cardboard prototypes, as this stage provided a balance between abstraction and concreteness while also preserving diversity of ideas.
Similar to \cite{moharana2019robots}, we clustered the visual and annotated information from these artifacts and tables into six themes (four robot roles across healthcare settings, robot appearance and embodiment, and robot interaction modalities).
The first author conducted the initial theme development, which the other two authors reviewed and discussed until full consensus was reached.

\textbf{2) RQ2–RQ4.} We coded transcripts from the interviews and post-education session discussions using thematic analysis \cite{clarke2017thematic} to examine participants' evolving perspectives on multidisciplinary collaboration (RQ2), their iterative refinement of ideas toward higher-fidelity prototypes (RQ3), and their learning curve and reflection on the educational sessions (RQ4). 
The first author developed the initial codes through iterative review of transcripts. The second author independently coded a subset of the data. 
Following collaborative discussions, a shared codebook was developed. 
Both authors then applied this codebook to the full dataset, and inter-coder agreement was assessed using Cohen's Kappa. 
The resulting codes were collaboratively synthesized into higher-level themes. 
For RQ2, the analysis resulted in two themes with two sub-themes ($\kappa = \,$0.91); for RQ3, two themes with three sub-themes ($\kappa = \,$0.94); and for RQ4, three themes with five sub-themes ($\kappa = \,$0.71). 
All values indicate substantial to almost perfect agreement.

\section{Results}

\begin{figure*}[!htbp]
    \centering
    \includegraphics[width=0.82\textwidth]{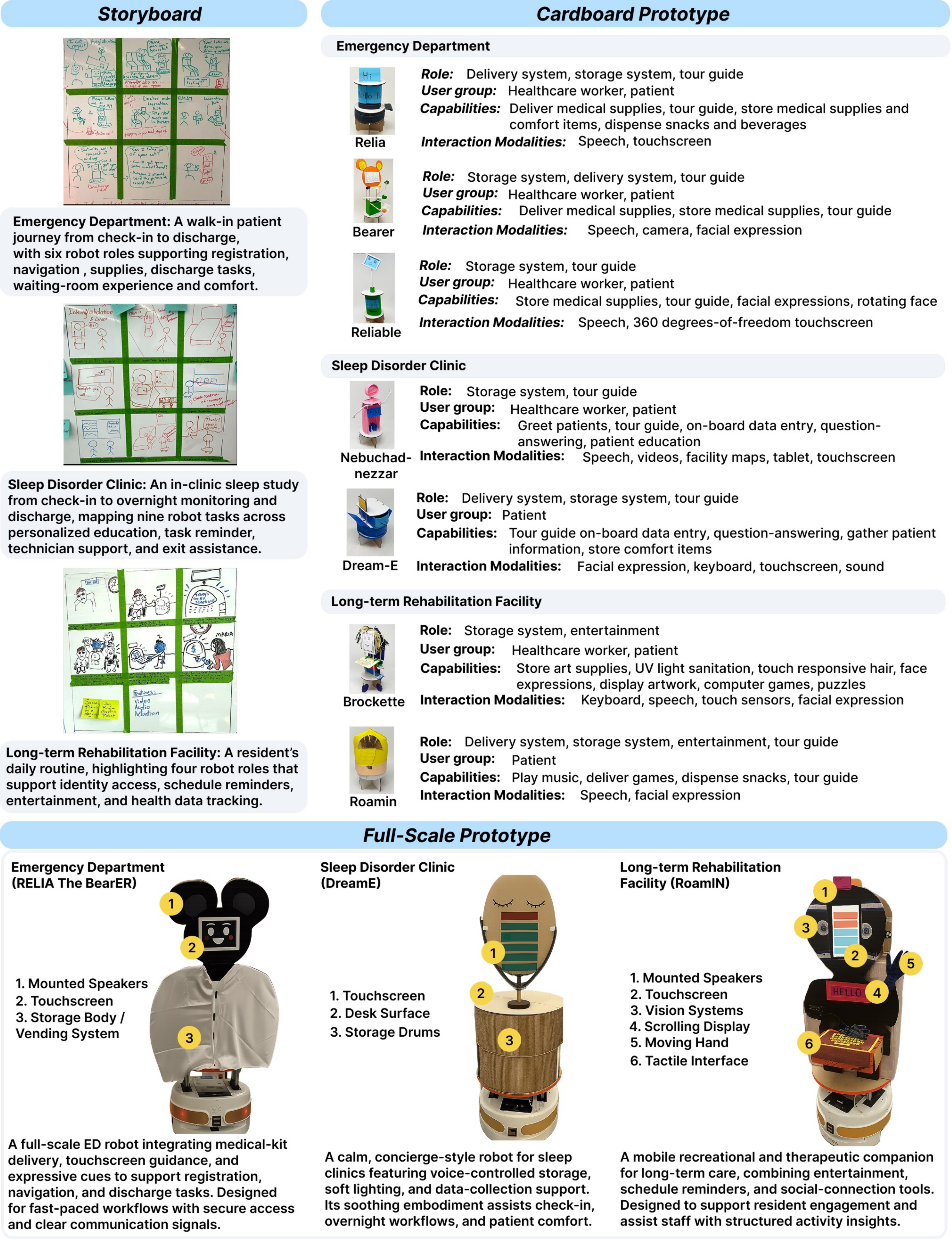}
    \caption{
        Evolution of co-ideation across three stages: 
        storyboards illustrating robot-assisted scenarios in ED (top), SDC (middle), and LTR (bottom);
        $\rightarrow$ Cardboard prototypes elaborating each context's envisioned robot roles, user groups, capabilities, and interaction modalities for ED (top), SDC (middle), and LTR (bottom);
        $\rightarrow$ full-scale prototypes developed by participants for ED (left), SDC (middle), and LTR (right). Comparison of robot roles across healthcare settings:  1) Delivery roles were envisioned in ED, SDC and LTR (shared, with distinct purposes); 2) Storage roles appeared in ED, SDC and LTR (shared, with distinct purposes); 3) Tour guide roles emerged in ED, SDC and LTR (shared but contextually distinct with different levels of emphasis); 4) Comfort/entertainment roles were unique to LTR.}
        \Description{This figure shows the evolution of co-ideation artifacts across three healthcare settings: Emergency Department (ED), Sleep Disorder Clinic (SDC), and Long-Term Rehabilitation (LTR). For each setting, the figure displays three rows of artifacts: (1) early storyboards illustrating envisioned robot-assisted scenarios; (2) cardboard prototypes depicting robot roles, user groups, capabilities, and interaction modalities; and (3) full-scale prototypes created by participants. A comparison highlights differences and overlaps in robot roles across settings: delivery and storage roles appear in LTR, ED, and SDC with different purposes; tour guide roles appear in LTR, ED, and SDC but are contextually distinct with different levels of emphasis; and comfort or entertainment roles are unique to LTR.
        }
    \label{fig:artifacts}
\end{figure*}

\subsection{RQ1: Cross-Contextual Analysis of Co-Designed Artifacts}
\label{sec:linkto5}

We present artifacts from each design stage and examine how they reflect both shared and context-specific insights across the three healthcare settings.

\subsubsection{Design Artifacts Across Stages}

Participants created a total of \textbf{19 storyboarded robot roles} across the three healthcare settings:
ED featured 6 roles, ranging from registering and guiding patients, to offering supplies, providing comfort, and supporting discharge processes. 
LTR included 4 roles, focusing on detecting patient identity, retrieving schedules, delivering entertainment, and tracking therapeutic interactions. 
SDC explored 9 roles, including greeting and orienting patients, providing personalized education, collecting nighttime feedback, supporting technicians, managing data, and assisting with exit procedures.

In the subsequent cardboard prototyping phase, teams constructed \textbf{a total of seven cardboard prototypes} across the three healthcare settings (ED: 3, SDC: 2, LTR: 2).
Each prototype elaborated on intended user groups, task-specific features (e.g., sanitation, delivery, data entry), and interaction modalities such as speech, facial expression, touchscreen monitors, and physical input devices.

Each team then advanced to constructing \textbf{one full-scale prototype per setting} (three in total). 
These final prototypes synthesized selected ideas from the earlier storyboards and cardboard prototypes, consolidating the most prioritized roles, features, and interaction modalities into full-scale embodiments.
These included: 
1) \textit{RELIA the BearER} (ED): A medical kit–dispensing robot for emergency departments. Equipped with a patient-facing display, barcode verification, and a vending-machine-like body, it aimed to streamline discharge and kit retrieval.
2) \textit{DreamE} (SDC): A sleep clinic concierge and relaxation assistant. With calming lights, a mirrored touchscreen, and voice interaction, DreamE created a peaceful care experience and assisted staff with delivery tasks.
3) \textit{RoamIN} (LTR): A multifunctional therapeutic assistant for long-term care, focused on entertainment, schedule display, and family connection. It featured components such as mounted speakers, a touchscreen, a facial recognition camera, and a tactile text interface. 

Figure \ref{fig:artifacts} shows an overview of artifacts developed throughout the co-design. 
Details on storyboard descriptions, prototype component annotations and functionality can be found in Appendix \ref{app:activities_detail}.

\subsubsection{Comparing Robot Roles Across Healthcare Facilities}

Based on the findings shown in Figure \ref{fig:artifacts}, we categorized the prototypes into four themes of robot use-cases: (1) robots as delivery systems, (2) robots as storage systems, (3) robots as tour guides, (4) robots for comfort and entertainment. 
These themes are further categorized into robots that support patients (i.e., patient-focused) and those that assist HCWs (i.e., HCW-focused) in the workplace.
Our analysis revealed shared needs across facilities, alongside distinct emphases shaped by differing patient populations and service contexts.

\textbf{Robots as a delivery system (ED, SDC and LTR).} The themes of delivery and storage are closely linked, as all participant-designed robots feature mobile bases, and delivery tasks inherently require storage. 
For clarity, this theme focuses on delivery, where timely transportation is essential to prevent workflow delays.
Both the ED and SDC groups envisioned patient- and HCW-focused robots with delivery roles. 
In ED, delivery was often critical due to the urgency of patient care.
For example, the HCW-focused Bearer-ED, was designed to fetch medical kits directly to the room before the doctor arrives, ensuring seamless treatment: \textit{``When patients are in the screening room, that robot can get the kit [ordered by the doctor] and bring it to the room that is required. The doctor can [then] meet the patient and have the kit ready in that room at the same time'' (P1)}. 
For SDC, the HCW-focused Dream-E-SDC were designed to transport supplies such as electrodes or masks to technicians, saving them the effort of \textit{``running back to the storage room (P2)''}.
Dream-E-SDC was equipped with an enlarged storage compartment to carry essential equipment and bulkier supplies.

\textbf{Robot role as a storage system (ED, SDC and LTR).} While all three facilities incorporated some form of storage in their robot designs, storage-focused support emerged most clearly in the LTR setting, integrated into patients' daily routines.
For instance, the bottom section of Brockette-LTR, an art-therapy robot, included three sections for \textit{``supplies, games, or whatever is necessary'' (P6)}, offering flexible and customized support.
In contrast, in ED and SDC, the storage compartments on robots like Relia-ED or Dream-E-SDC were not standalone features, but functionally tied to their delivery responsibilities to carry medical kits, test supplies, or data-entry devices in support of fast-paced, HCW-focused tasks.
For example, Relia-ED could respond to a HCW's request to pass along an item: \textit{``The clinician in that room would say, `Oh, could you please deliver this to another clinician?' Then the storage unit just opens up, closes, and goes to the other clinician, notifies them, `I've got a package delivered for you,' and it opens up.'' (P21)}

\textbf{Robot role as a tour guide (ED, SDC and LTR).} The role of patient-focused robots as tour guides was most salient in the SDC, where patient onboarding posed a repetitive burden on HCWs. 
Because patients arrived in staggered shifts and required detailed orientation, \textit{``Technicians are interrupted because they have to go out, greet patients, verify who's there, and then explain what's going to happen over and over again. The technicians are already performing very robotic tasks'' (P3)}. 
To reduce this burden, Nebuchadnezzar-SDC is expected to \textit{``get patients acclimated and accustomed to where they are, as well as the safety features and the upcoming uncertainty about what they're doing'' (P3}).
In the ED, robots like Relia-ED also served as patient guides, but with a more urgent and destination-focused orientation. Their primary function was to escort patients efficiently through clinical spaces, such as \textit{``to the next room after the waiting room'' (P21)}, minimizing delays in care pathways.
By contrast, in the LTR setting, where patients were already familiar with their environment, the tour guide role was less emphasized. 
Although robots like Roamin-LTR could ``escort'' patients, their primary function was recreational and therapeutic, rather than structured spatial orientation.

\textbf{Robot roles for comfort and entertainment (LTR).}
Only participants in the LTR group emphasized the robot's patient-focused role in entertainment and social interaction. 
The facility was described as the equivalent of \textit{``11 football fields''} with over 500 residents, so HCWs do not have enough time to provide all these interactions.
For example, Roamin-LTR was designed to assist with recreational activities like dispensing snacks and playing music, because \textit{``all the staff are very busy with medical stuff or record-keeping, while the patients really want to socialize.'' (P5)}. 
Some participants also envisioned integrating robots into therapeutic contexts, such as art therapy, where data tracking on leisure activities might help monitor patient mental states, an idea that emerged during the storyboarding phase: \textit{``The art therapist is getting a receipt of interactions from the robot at the end of the day.'' (LTR storyboard annotation)}.
\subsubsection{Robot Embodiment and Interaction Modalities Across Healthcare Facilities}

\textbf{Robot embodiment: friendly appearance and accessible interaction.} Across \textit{all settings}, participants emphasized friendly and welcoming robot appearances to promote interaction comfort and social acceptance.
For ED, Bearer-ED had a teddy bear-like appearance to provide a \textit{``cute embodiment that the patients would feel comfortable with'' (P1)}.
P21 also highlighted the 360-degree screen on Reliable-ED that displays messages from any angle, ensuring patients do not miss critical information regardless of their position.
For LTR, Brockette-LTR included interactive features like touch-responsive ``hair'',  designed with a plushy, soft texture to evoke comfort and emotional reassurance.
When touched, the ``hair'' would move or light up, accompanied by friendly verbal prompts such as \textit{``Hi, how are you today?'' (P6)}. 
To ensure hygiene, LTR group also suggested using UV lights to sanitize the surface after each use.
For SDC, Nebuchadnezzar-SDC featured a smiley expression to \textit{``break the ice'' (P3)}, saying ``follow me'', as it guided them on a facility tour.

\textbf{Robot interaction modalities.} Among the interaction modalities explicitly noted by the creators of the seven cardboard prototypes, speech was the most frequently mentioned modality (6/7), followed by facial expressions, and touchscreens (both 4/7). 
Additional modalities included sound, cameras, facility maps, tablets, educational video streaming, and keypads.
Speech was widely utilized for its versatility and necessity across tasks, enabling seamless communication with the users. 
Facial expressions, emphasized by at least one prototype in each facility, played a key role in fostering social engagement and enhancing the robot's approachability and acceptance. 
Touchscreens offered a convenient interface for tasks requiring user input, such as when Dream-E-SDC facilitated questionnaire feedback from patients or allowed HCWs to operate robotic embodied equipment like drawers.
Beyond these commonly used modalities, unique features catered to specific scenarios. 
For instance, Nebuchadnezzar-SDC incorporated facility maps and videos for customized education purposes to provide an informative guided tour and educational content. 
The keypad on Brockette-LTR served an entertainment-focused purpose (e.g., onboard games or puzzles).

\begin{table*}[t]
\caption{Themes and sub-themes identified across three research questions (RQ2–RQ4). 
For design process evolution, the workshop consisted of three iterative stages: 
storyboard $\to$ cardboard prototype $\to$ full-scale prototype (referred to as stage 1 $\to$ stage 2 $\to$ stage 3 in the table for simplicity, indicating when sub-themes emerged most prominently).}
\label{tab:results}
\Description{Each theme captures a major focus area emerging from the co-design workshop analysis, with sub-themes representing finer-grained patterns.}

\centering
\begin{tabular}{p{2.9cm} p{6.35cm} p{7.25cm}}
\toprule
\textbf{Dimension} & \textbf{Theme} & \textbf{Sub-theme} \\
\midrule

\makecell[l]{\textbf{Multidisciplinary} \\ \textbf{Collaboration} (RQ2)} & 
\makecell[l]{1) Sustained Collaborative Learning \\\hspace{0.75em} Refines Solutions and Builds Confidence} &
\makecell[l]{1.1) Cross-Expertise Learning: \\\hspace{1.75em}Informing Design and Building Confidence \\ 
1.2) Cross-Team Knowledge Exchange: \\\hspace{1.75em}Inspiring and Refining Solutions} \\
\cmidrule(l){2-3}

& \makecell[l]{2) Multifaceted Identities and Expanded \\\hspace{0.75em} Contributions} &
\\
\midrule

\makecell[l]{\textbf{Design Process} \\ \textbf{Evolution} (RQ3)} &
\makecell[l]{1) From Sketch to High-Fidelity Prototypes:\\\hspace{0.75em}  Iterative Transitions in Embodiment and \\\hspace{0.75em} Envisioned Use Cases} &
\makecell[l]{1.1) Designing for Welcoming Appearance and Tactility \\\hspace{1.75em}(Stage 1 $\to$ Stage 2)\\
1.2) Adapting Designs for Ergonomic and Spatial Fit \\\hspace{1.75em}(Stage 2 $\to$ Stage 3)\\
1.3) Deployment Planning and Anticipatory Functionality \\\hspace{1.75em}(Stage 2 $\to$ Stage 3)} \\
\cmidrule(l){2-3}

& 
\makecell[l]{2) From Exploration to Focus: \\\hspace{0.75em} Design Prioritization Through Iteration and \\\hspace{0.75em} Real-World Constraint Consideration \\\hspace{0.75em} (Stage 1 $\to$ Stage 3)} 
& 
\\
\midrule

\makecell[l]{\textbf{Educational} \\\textbf{Sessions} (RQ4)} &
\makecell[l]{1) Co-Learning Robot Design Considerations:\\\hspace{0.75em} Exploring Trade-Offs and Human–Robot \\\hspace{0.75em} Collaboration} &
\makecell[l]{1.1) From Imaginative Concepts to Constraint-Aware \\\hspace{1.75em}Trade-Offs\\
1.2) Human–Robot Collaboration} \\
\cmidrule(l){2-3}

& 
\makecell[l]{2) Emerging Engineering-Oriented Reasoning:\\ \hspace{0.75em} From Motion Modules to Real-World Challenges} &
\\
\cmidrule(l){2-3}

&
\makecell[l]{3) Eliciting Real-Life Connections:\\\hspace{0.75em} Integrating Insights from Daily Life and Work} &
\makecell[l]{3.1) Lived Awareness of Spatial Constraints\\
3.2) Trade-Offs Between Safety and Functionality\\
3.3) Professional Lenses Prompt Critical Reflection} \\

\bottomrule
\end{tabular}
\end{table*}

\subsection{RQ2: Influence of Multidisciplinary Collaboration on Co-Design Dynamics}
This section shows how close, sustained multidisciplinary collaboration fostered ongoing knowledge exchange and perspective-sharing, resulting in richer and more collective contributions that extended beyond professional titles and stakeholder labels.

\subsubsection{1) Sustained Collaborative Learning Refines Solutions and Builds Confidence}

The interviews revealed two key dimensions of sustained collaborative learning: \textit{exchanges across diverse expertise} and \textit{idea sharing across teams}.

\textbf{Cross-Expertise Learning: Informing Design and Building Confidence. } 
Eight of eleven interviewees highlighted that the close collaboration fostered knowledge exchanges in which they continuously learned from one another, which not only grounded the co-design process in practical knowledge but also strengthened participants' confidence in contributing their own ideas.
Participants gained concrete insights into healthcare workflows and professional practices through interactive activities such as the stakeholder panel and dollhouse exercise: \textit{``What I never knew before was the doctor's point of view, what they would be doing during that day.'' (P21)}
At the same time, HCWs valued the perspectives of artists, engineers, makers, patients, and caregivers. 
As P17 noted: \textit{``I've talked to patients for decades about their experiences in healthcare. There is always something from everybody's experience that people bring to an interaction, and in a group of people with different life experiences and different roles, you can always learn something from everybody.''}
This collaborative learning also contributed to participants' confidence to contribute when they found their ideas aligned with those of professionals.
P6 explained: \textit{``It made me feel that my ideas are not only creative but also useful, given the professional level at which these people work.''} 

\textbf{Cross-Team Knowledge Exchange: Inspiring and Refining Solutions. } Two of the eleven interviewees specifically emphasized how seeing others' creative robot ideas, in terms of functions or appearance, sparked inspiration and improvements in their own work.
P6 valued the weekly progress review and visual documentation (e.g. photos and recordings), which allowed her to reflect on her own progress and draw inspiration from others' designs, expressing excitement about \textit{``try[ing] these new ideas.''}
Similarly, P9, an artist, who transitioned from the LTR group to help the SDC group, highlighted how the circulation of people between groups helped spark new ideas: \textit{``P6 went to a different group and brought a different perspective. Other artists also brought different perspectives to different groups. Now I'm working with the SDC group, but I still want to talk to the LTR group.''}

\subsubsection{2) Multifaceted Identities and Expanded Contributions}

In-depth interviews revealed that 6 of 11 participants contributed through \textit{multifaceted identities} that extended beyond their primary professional titles (e.g., HCWs, engineers, artists) or stakeholder labels (e.g., caregivers, patients). 
These combined perspectives enriched the co-design process by enabling participants to bridge different viewpoints. For instance, P8, a resident in an LTR facility, shared that he is also a licensed architect and an artist,  enabling him to both convey resident needs and contribute to prototyping.
Similarly, P6, an artist, drew on her \textit{approximately 18 years} of caregiving experience with individuals with developmental disabilities, while P5 combined her profession as an artist and former recreational therapist with her lived experience as the caregiver of a chronically ill partner:  \textit{``I had a life partner who was chronically ill, so I spent a lot of time sitting in emergency rooms when I wasn't sick, just to observe what was going on there. So what I have to contribute is multifaceted.''}
We found that some participants, though formally labeled as a specific stakeholder group (e.g., HCWs), contributed perspectives extending far beyond that identity.
For example, HCWs had internalized patients' and caregivers' viewpoints alongside their professional expertise, enabling them to offer more integrative and multifaceted insights. 
As P17, an HCW with decades of experience in treating patients, explained: \textit{``You need to understand human nature and how patients feel. Ten people in the same situation will have ten different lived experiences---what they think was good, bad, missing, miscommunicated, hurtful, or microaggressive. You need an open mind, open ears, and open heart to listen and bring those perspectives into your care.''}

\subsection{RQ3: Iteration as a Pathway to Refinement and Real-World Relevance}

This section examines how iterative, hands-on co-ideation activities helped participants transform abstract ideas into increasingly concrete, high-fidelity robotic prototypes. 
As the fidelity of robot prototypes increased, they refined appearance and tactility to enhance stakeholder acceptance, clarified envisioned use cases, and, as construction effort increased, began prioritizing core design needs.

\subsubsection{1) From Sketch to High-Fidelity Prototypes: Iterative Transitions in Embodiment and Envisioned Use Cases}

\textbf{Designing for Welcoming Appearance and Tactility (Storyboard to Cardboard Prototype). } 
As participants transitioned from storyboards to cardboard prototypes, they began focusing on making robots appear friendly and welcoming rather than intimidating. 
Many refinements emerged only once abstract ideas were embodied: \textit{``You have something in mind, but when you start putting it together, it doesn't feel quite right. So we adjusted some things a bit, but I think we kept most of the storyboard and its integrity.'' (P19)}
These adjustments often centered on aesthetics and tactility: In the LTR team, P19 noted, \textit{``We wanted to make it not frightening, but welcoming.''} 
The ED team similarly moved from a basic utilitarian form to a more inviting one, as P1 recalled: \textit{``We started with just a head and a base. Then someone suggested making it more friendly, not just furniture. So we rounded it out to look more like a bear.''}

\textbf{Adapting Designs for Ergonomic and Spatial Fit (Cardboard Prototype to Full-Scale Prototype). } 
As prototypes scaled up from cardboard to full-size, participants adjusted their designs to address spatial and ergonomic fit. 
In the LTR team, P19 recalled, \textit{``We aimed a little too small before. Then, when putting it together, we realized it could be more forgiving, so we added inner space and made the drawer bigger.''} 
The SDC team similarly modified their prototype for ease of use, with P2 noting, \textit{``We swapped the order of the drawers. This way, it's at eye level and the technician doesn't have to stoop.''}

\textbf{Deployment Planning and Anticipatory Functionality (Cardboard Prototype to Full-Scale Prototype). } 
At the full-scale stage, participants looked beyond the prototype, anticipating integration needs and imagining extended functionalities.
In the SDC team, P2 emphasized infrastructure requirements: \textit{``The electronics would need a power source for the iPad.''} 
They also looked ahead to expanded functionality. As P2 explained, \textit{``Functionality was our biggest issue. We wanted a movable storage for the technologist and an information source for patients. Eventually, we hope to integrate with the Electronic Medical Records (EMR) to allow data interface.''}
Workflow alignment was another focus. P2 explained that the robot should \textit{``stock everything neatly''} and \textit{``have some interface to educate patients''}, allowing technicians to focus on tasks \textit{``computers can't do''.}

\subsubsection{2) From Exploration to Focus: Prioritizing Designs Under Constraints (Throughout the Workshop)}
As prototypes advanced from sketches to cardboard and full-scale models, participants encountered increasing demands in time, coordination, and technical effort. 
These escalating challenges led them to treat the process as analogous to real-world development pressures, prompting difficult choices about scope and priorities.

In the LTR team, P19 explained how limited resources forced a sharper focus: \textit{``We had fewer resources, so we had to focus on one thing.''} 
A HCW from the ED team similarly noted that they evaluated which designs were worth pursuing under resource constraints, reflecting, \textit{``We'd be more likely to spend the money on [certain functions].'' (P17)} 
Another ED team member, P1, described how early brainstorming produced many ideas, but prioritization narrowed them to what clinicians found most useful: \textit{``We started with many different ideas, and then we focused on just one function, based on what the doctor said would be most useful.''} 
This led the team to emphasize a supply-delivery robot while setting aside options such as a check-in assistant temporarily.

Prioritization also involved consolidating overlapping concepts that had emerged during early brainstorming. 
As P19 from the LTR team recalled, \textit{``We were together, then spread out. Similar ideas were born from the storyboard.''} 
Consolidation helped teams streamline redundant ideas and concentrate their limited time and effort on more viable design directions.

\subsection{RQ4: Structuring Educational Sessions for Inclusive Technical Engagement}

This section illustrates how educational sessions supported participants with less technical experience, enabling them to engage in engineering reasoning through co-learning around design trade-offs and linking abstract concepts to real-life experiences.

\subsubsection{1) Co-Learning Robot Design Considerations: Exploring Trade-Offs and Human–Robot Collaboration}
Participants engaged in lively discussions about how robots should function across real-world contexts.
Rather than pushing toward binary judgments (e.g., whether a design was feasible), the sessions invited participants to explore what makes a robot useful, appropriate, or collaborative depending on different use cases.

\textbf{From Imaginative Concepts to Constraint-Aware Trade-Offs.}
Through guided dialogue, participants' initial proposals often reflected unconstrained creativity, for instance, imagining a \textit{``flying angel'' (P12)} robot or one that \textit{``lives on a track on the ceiling'' (P17)}. These ideas highlighted the openness of brainstorming but gave little attention to technical or environmental constraints.
As researchers, we used these moments to scaffold discussions about design trade-offs, illustrating how efficiency, safety, and task goals interact through examples such as robotic arms in pick-and-place tasks \cite{lyu2024farpls}. 
As conversations progressed, participants began to incorporate constraint-aware reasoning, for instance, suggesting that the robot should \textit{``find a backup route'' (P17)} when facing obstacles. 
Importantly, these exchanges were not one-sided lectures: participants also built on each other's ideas in real time. 
For example, P8 critiqued the \textit{``flying angel''} proposal by remarking, \textit{``The noise is unbearable''}, reframing the idea in terms of technical feasibility.

\textbf{Human–Robot Collaboration. } Rather than viewing robots as replacements, participants were encouraged to see them as collaborators, leveraging robot precision and endurance alongside human adaptability and judgment \cite{weidemann2023literature}. 
This human-robot collaboration perspective became concrete in discussions about delivery and supply distribution, where participants envisioned robots handling routine logistics while humans provided oversight and flexibility.
P6 described the value of assigning robots simple, repeatable behaviors: \textit{``The robot is programmed, for example, to stop every 35 feet or every 15 feet. Then it continues on. That seems to be a reasonable set of commands and abilities for a robot.''} 
At the same time, P9 emphasized lightweight but essential human involvement: \textit{``Supplies are already here, and [a human] can pass [the supplies] out as the robot comes.''}
Finally, P6 highlighted the flexibility of human presence, noting: \textit{``[Suppose] the trays are there, `Oh, no, I don't want any applesauce', then the staff don't need to put it on. It's very simple.''}
Together, these reflections illustrate a transition from imagining fully autonomous systems toward envisioning collaborations in which robots handle routine logistical tasks while humans provide flexibility and contextual judgment.
\subsubsection{2) Emerging Engineering-Oriented Reasoning: From Motion Modules to Real-World Challenges.}

While earlier themes focused on design trade-offs and context-sensitive appropriateness, this theme highlights how participants began to think about the underlying technical mechanisms that enable robot behaviors. 
Despite coming from non-technical backgrounds, many participants demonstrated an emerging ability to reason in structured, engineering-oriented ways---considering how robot functions might be achieved, reused, or adapted for different use cases.
At a functional level, they imagined how robot motions could be repurposed for other tasks. 
For example: \textit{``If it has an up-and-down movement, maybe it could also do a side-to-side motion.'' (P12)} P17 likewise speculated on reusing a drawer's push-pull action: \textit{``If the drawer goes in and out, you could theoretically use it to push a button on an automatic door.''}

At a broader system level, participants extended their thinking to anticipate real-world robotic challenges beyond the educational content. 
For example, after a discussion on obstacle avoidance, P5 noted spatial limitations that could hinder a robot's movement: \textit{``The robot stops, backs up, and then turns, but there can be no place to turn. It might end up going into someone's room or somewhere it's not supposed to go.''} 
Others envisioned how sensor feedback could influence decision-making: \textit{``Have a sensor that says a person is coming. Then if I'm the robot, I should stop so the person can go by.'' (P6)} or \textit{``Imagine a sensor says: nobody here.'' (P11)}
Regarding mapping problems, although our demonstration showed the robot operating in a static, uncluttered lab, participants anticipated challenges in more dynamic environments. 
P8 asked: \textit{``What happens if I move a bunch of stools around? Does the map refresh that quickly?''} and P2 raised concerns about mismatches between robot scale and map resolution: \textit{``If you're only mapping the area up to 1 foot, and we're now building a 4-foot robot, so it could still bump into things and stick out higher.''}
Although participants did not use technical terminology, their reflections revealed a grasp of core robotics challenges, such as environment-aware navigation and map updating, that align with concerns in research on 2D-LiDAR (Light Detection and Ranging) -based SLAM (Simultaneous Localization and Mapping) \cite{ye2021mapping} and Dynamic Pose Graph SLAM techniques \cite{walcott2012dynamic}.

\subsubsection{3) Eliciting Real-Life Connections: Integrating Insights from Daily Life and Work}

Educational sessions also elicited non-technical participants to link robot design to their daily lives and professional practices. 
These lived reflections helped them better grasp design challenges while broadening perspectives on robot design.

\textbf{Lived Awareness of Spatial Constraints. } Non-technical participants drew from everyday experiences to highlight how robots must adapt to human environments. 
P5, an artist, recalled an \textit{``instant dislike''} of a robot on campus that blocked a narrow sidewalk, forcing them to step onto the grass: \textit{``The environment and the robot have to match and work together.''}
Others drew from their observations in healthcare, highlighting how ED spaces dynamically shrink under pressure: \textit{``The aisles become narrow when it gets busy and gurneys line the halls'' (P5)}, and \textit{``Sometimes even the nurses are walking sideways'' (P12)}.

\textbf{Trade-Off Between Safety and Functionality. } Participants with fewer technical backgrounds translated personal observations into reflections on the trade-off between safety and functionality in real-world environments. 
P8, an artist and living in the LTR facility, recalled a conversation with a technician who explained why certain machines require human supervision due to safety risks: \textit{``There are moving parts, and if someone steps up and it catches their limb, that's a real safety risk.''} 
P8 transferred this insight to the design of healthcare robots, reflecting: \textit{``From a healthcare standpoint… maybe the fewer moving parts [of the robot's mechanisms], the better. Even a simple back-and-forth motion [such as a drawer sliding] can be a concern.''} 
This demonstrates how personal insights were transferred into design logic, reinforcing the need for safer robot motion patterns. 
From another angle, P5 referenced the experience of a friend taking autonomous taxis, sharing concerns about over-engineered safety leading to reduced utility: \textit{``The taxis detect obstacles and just stop there... people are sitting in the cabs and they're not moving''(P5)}.

\textbf{Professional Lenses Prompt Critical Reflection. } 
During the educational session, we demonstrated a robot in a controlled lab environment, but participants quickly noted that real facilities also include wet spaces such as showers.
P2, who works in SDC, drew on her professional experience to raise overlooked challenges. 
She pointed out that their facility includes open bedrooms and bathrooms, as well as wet shower areas used by disabled individuals, and questioned whether the robot could function reliably in such settings: \textit{``Does the robot have water detachment? What if it goes into a surface that is watered?'' (P2)} 
Rather than passively accepting the demonstration, P2 actively applied her professional lens to reveal environmental challenges the team might otherwise have missed.

\section{Discussion}

\subsection{Summary of Findings}

Through a 14-week multidisciplinary co-design workshop across the ED, SDC, and LTR settings, we examined how participants envisioned robotic solutions across healthcare contexts (RQ1), and how multidisciplinary collaboration (RQ2), iterative design (RQ3), and educational components (RQ4) shaped the co-creation process.

For RQ1, participants' envisioned robot roles and interaction modalities showed clear cross-context commonalities yet diverged according to facility needs. 
Across ED, SDC, and LTR, most teams emphasized that robots should be friendly, approachable, and easy to interact with, primarily through speech-based interaction, which was mentioned across nearly all prototypes, along with facial expressions and touchscreen interfaces.
However, the characteristics of each setting shaped distinct role priorities; for example, the ED and SDC prototypes focused on delivery to support fast-paced clinical tasks, while LTR robot design emphasized storage and comfort/entertainment to support residents' daily lives. 
Even shared roles were framed differently, as tour guides were envisioned for rapid navigation in ED versus onboarding and information in SDC. 
These patterns suggest that robot roles should align with facility-specific priorities while attending to contextual nuances.

For RQ2, we found that when participants collaborated closely within the same group, rather than contributing separately, their professional perspectives, personal experiences, and design ideas continuously interacted and cross-pollinated. 
This not only fostered knowledge sharing and mutual learning, but also built participants' confidence and engagement through iterative exchanges, creating a positively reinforcing collaborative climate.

For RQ3, we found that the progression from low- to high-fidelity prototypes involved not only increasing realism in appearance and tactility, but also an evolving in thinking. 
Early storyboards supported divergent brainstorming; mid-stage cardboard prototypes prompted consideration of how robots could be embodied and situated in real environments; and full-scale prototypes required participants to confront practical constraints such as scale, ergonomics, and future deployment. 
Together, these stages played complementary roles: low-fidelity stages support broad exploration, while high-fidelity stages drive integration and refinement.

For RQ4, we found that embedded educational sessions helped non-technical participants transition from imaginative brainstorming to more constraint-aware, engineering-oriented reasoning, while also prompting them to connect robot design to their everyday experiences and professional practices. 
This broadened their understanding of human–robot collaboration, design challenges, and technical feasibility, enabling more grounded design judgments. 
At the same time, these user-centered perspectives rooted in daily life and work experiences also enriched technical participants' understanding, further fostering bidirectional knowledge construction within the multidisciplinary teams.

\begin{table*}[!htbp]
\centering
\begingroup
\caption{Four embodied-AI-specific co-design dimensions and eight guidelines distilled from our 14-week, multi-context study. 
These guidelines articulate what makes co-design \textit{embodied}: grounding design in bodily routines and spatial realities; revealing feasibility through physical and social constraints; building stakeholders' embodied literacy over time; and expanding the design space toward deployment-oriented reasoning. Each guideline is paired with illustrative examples from the Emergency Department (ED), Long-Term Rehabilitation (LTR), and Sleep Disorder Clinic (SDC) contexts from our workshop series.}
\label{tab:research_guidelines}

\begin{tabular}{p{0.5cm} p{8.2cm} p{8cm}}
\toprule
\multicolumn{3}{l}{\textbf{Dimension 1: Embodied Needs Grounding}}\\
\midrule \textbf{Guideline} & &\textbf{Example} \\ 
\midrule

\textbf{G1.1} &
\textbf{Identify Embodied Needs Beyond What Disembodied Systems Cannot Fulfill}: 
Surface needs that inherently require embodied presence, spatial sensing, mobility, or co-located interaction, rather than defaulting to technical-hammer assumptions. &
e.g., HCWs from SDC envisioning roles (warm greeting, escorting, handing out surveys, answering questions, transporting items, checking patient status) that current stationary kiosks cannot adequately fulfill. \\

\textbf{G1.2} &
\textbf{Map Real Workflows and Situated Physical Contexts to Ground Design}:
Document physical setup, spatial constraints, environmental cues, and workflow patterns that shape where an embodied agent must fit. &
e.g., on-site facility visits; VR/photo walkthroughs of rooms and corridors; dollhouse activity mapping patient flow, staff movement, and real workflow sequences. \\

\midrule
\multicolumn{3}{l}{\textbf{Dimension 2: Embodied Constraints \& Feasibility}}\\
\midrule

\textbf{G2.1} &
\textbf{Reveal Constraints Through Iteration Toward Higher-Fidelity Prototypes}:
Progressively transition from low- to higher-fidelity embodiments to surface appearance, tactility, spatial fit, ergonomic, and interaction-coordination factors that only emerge when ideas are instantiated physically. &
e.g., storyboard → cardboard → full-scale prompting reflections on non-intimidating appearance, tactile comfort, human-robot collaboration in LTR, and adjusting drawer height for ergonomic bending in SDC. \\

\textbf{G2.2} &
\textbf{Reflect on Feasibility Constraints to Guide Convergent Prioritization}:
Highlight decisive feasibility factors (e.g., engineering effort, cost, deployment practicality) to help participants balance desired features with realistic resource limits. &
e.g., ED team prioritizing rapid tool-transfer and patient communication as high-impact for acute care after recognizing engineering effort required for high-fidelity refinement (used as an analogy for real-world deployment cost considerations). \\

\midrule
\multicolumn{3}{l}{\textbf{Dimension 3: Embodied Literacy Building}}\\
\midrule

\textbf{G3.1} &
\textbf{Bridge Abstract Concepts via Concrete Demonstrations and Lived-Experience Reflections}:
Ground technical concepts in tangible robot behaviors and create discussion space for participants to relate them to daily work/life insights. &
e.g., live demos and scenario clips; prompting reflections such as robots avoiding personal-space intrusion after participants shared stories of being edged out of narrow walkways. \\

\textbf{G3.2} &
\textbf{Tailor Technical Depth to Fidelity Stage and Evolving Understanding}:
Increase technical detail progressively while using feedback and reflections to adapt content to what resonates with participants. &
e.g., participant-requested robot history during interviews; introducing navigation, obstacle avoidance, sensing limits, and autonomy during full-scale prototyping. \\

\midrule
\multicolumn{3}{l}{\textbf{Dimension 4: Embodied Design Space Expansion}}\\
\midrule

\textbf{G4.1} &
\textbf{Integrate Diverse Embodied Sensibilities to Enlarge the Design Space}:
Combine artists', HCWs', caregivers', engineers', and patients' embodied intuitions to surface opportunities no single perspective reveals alone. &
e.g., artists highlighting material warmth; HCWs emphasizing workflow fit; engineers refining stability and sensing; caregivers noting calming cues. \\

\textbf{G4.2} &
\textbf{Adopt a Strengths-Based Lens to Expand Creative Contribution}:
Invite participants to share lived experiences, skills, and personal stories to shift from need-only framing toward value-adding co-creation. &
e.g., interviews surfacing participants' hidden skills and contributions, inspiring new co-creation directions. \\

\bottomrule
\end{tabular}
\endgroup
\end{table*}

\subsection{Transferable Guidelines from Long-Term Co-Design Workshops on Healthcare Embodied AI}

Our analysis distills fourteen weeks of multi-context workshops into a four-dimensional framework (eight guidelines in total) that collectively characterizes what makes co-design \textit{embodied} and identifies the design scaffolds required for embodied AI (See Table \ref{tab:research_guidelines}). These dimensions highlight:
(1) grounding design in lived, embodied needs;
(2) refining ideas through reflection on physical and social constraints revealed as fidelity increases;
(3) building participants' embodied literacy to enable more informed co-ideation; and
(4) expanding the design space by integrating diverse embodied sensibilities across multidisciplinary teams.

\paragraph{Dimension 1: Embodied Needs Grounding.}

\textit{(G1.1)} Co-design for embodied AI should begin from problems and needs, rather than from available technical capabilities (i.e., a ``technical hammer'' approach). For example, P2 from SDC indicated that their current stationary kiosk is not sufficient for greeting and guiding patients around the facility during onboarding tours, and for supporting communication and updates, and that they need a more friendly appearance and richer functions captured in the robot's embodiment. \textit{(G1.2)}  We also found it crucial to understand how such needs sit inside real environments and workflows: site visits, walkthroughs, and dollhouse activities produced a shared understanding of who moves where, when, and why.
Without this grounding, design can easily misalign with actual spatial and workflow conditions, as prior work shows. For instance, in \cite{ren2024working}, telepresence robots in institutional dementia care were hindered by physical constraints in the real world (e.g., limited space, mobility patterns), leading frontline staff to report that robots were easily knocked over and difficult to integrate into daily routines.
Thus, we recommend grounding early design not just in abstract goals, but in embodied needs that must fit concrete spatial and workflow patterns.

\paragraph{Dimension 2: Embodied Constraints and Feasibility.}

 \textit{(G2.1)} As ideas moved from low-fidelity sketches to cardboard forms and then to full-scale models, participants began noticing issues that only became visible when designs were embodied, such as appearance, tactility, spatial fit, ergonomic bending, and the division of work between humans and robots. Prior work also shows how insufficient embodiment obscures critical feasibility issues. For example Ren et al. \cite{ren2024working} described how uncustomized robot designs failed to meet staff and residents' communication needs, with details such as screens being too small and robot speech volumes being too low for people with visual or hearing impairments. In our case, similar specific constraints emerged as participants became aware of details such as drawer height in SDC or whether screens require protectors in LTR. This awareness directly drove design refinement, illustrating how iterative embodiments can surface details that would otherwise be overlooked.
\textit{(G2.2)} Similarly, progressing toward higher fidelity prototypes made resource limits more salient. In other words, participants became more aware that building and maintaining a robot requires engineering effort, time, and budget, not just sketching. This helped groups such as the ED team prioritize functions to pursue first when everything could not be implemented at once. 
In this sense, we recommend iterative embodiment design, which not only makes ideas more detailed but also enables constraints and priorities to emerge through reflection-driven refinement.

\paragraph{Dimension 3: Embodied Literacy Building.}

Bringing stakeholders with different technical backgrounds raises a key question: \textit{how can we support them in making informed design decisions about robots?} Our findings show that abstract explanations of sensing and control can be difficult for non-technical participants to follow, whereas concrete demonstrations, such as observing robots \textit{``open drawers or responding to voice commands''} (P17), made capabilities more tangible.
We also observed that despite efforts to simplify the robot-control demo for a non-technical audience by focusing on core functions such as sensor integration and spatial mapping, participants still found this to be ``too technical''. This discrepancy reveals a gap between design intentions and participants' actual reception of the introduced concept. In contrast, sessions that introduced the history of robots in an accessible, less-technical way resonated more. As P19 noted, \textit{``Seeing how far things have progressed opened our minds to what else could be done''.}
To support ongoing learning and engagement, we suggest \textit{(G3.1)} using concrete demonstrations and inspiring learning from reflection on lived experience. Also \textit{(G3.2)}, incorporating regular reflective practices (e.g., post-session debriefings, and informal check-ins) to gauge participant receptivity. Offering flexible session options that allow participants to choose their next exploration path can also foster a more adaptive and responsive experience. Ultimately, embodied literacy must evolve dynamically alongside fidelity increases, rather than being treated as a one-time instructional step.

\paragraph{Dimension 4: Embodied Design Space Expansion.}

\textit{(G4.1)} Our findings align with prior work emphasizing the value of multidisciplinary co-design in social robotics \cite{ostrowski2021long,jung2018robots}, and extend this perspective by showing how cross-role dialogue fosters integrated solutions grounded in both professional expertise and lived experiences. Rather than conducting separate interviews and synthesizing perspectives later, we emphasized sustained in-group collaboration to keep participants in stable, multidisciplinary teams over time. In these close-knit groups, knowledge does not emerge as static inputs to be collected, but through ongoing, dynamic co-creation in which participants continuously learn from and inspire one another.
\textit{(G4.2)} We also build on critiques of deficit-based framings in health-technology research \cite{harrington2020forgotten}, which often overlook the capabilities of older adults. Our findings extend this perspective by showing that participants from long-term care and senior communities, often positioned as groups who articulate needs, also contributed as artists, makers, and therapists, drawing on lived experience to inform design decisions. To support such contributions, we recommend providing platforms (e.g., storytelling) where participants can share not only the structural and personal challenges they face, but also the capabilities, creative practices, and lived-experience strengths. Personal stories illuminate diverse forms of expertise, disrupt stereotypes, and surface meaningful design insights.

\subsection{Expanding the Design Space for Healthcare Embodied AI}
Building on the design outcomes presented in Section \ref{sec:linkto5}, this section articulates the broader methodological novelty of our co-design approach. Our approach reveals a richer design space for healthcare embodied AI across three dimensions that prior work and commercial deployments often consider separately: (1) workflow-grounded contextualization, (2) capability-aware envisioning, and (3) cross-context learning across healthcare services.

\paragraph{1) Workflow-grounded contextualization.}
Rather than treating robots as standalone tools, our participants envisioned them as collaborators embedded in daily clinical routines and as part of patients' lived experiences. Grounding design in workflow allowed us to connect robot roles to concrete care experiences. For instance, SDC robots were envisioned with soft internal lighting that glowed like a nightlight and pulsed gently to create a calming, ``breathing'' effect. In the ED, pre-arrival toolkit delivery was emphasized to synchronize with urgent clinical tempo.
Existing systems, such as PARO for elderly care companionship \cite{hung2019benefits}, Moxi for tool transport to assist nurses \cite{moxi2024}, or SERVI for food delivery \cite{servi2025}, fulfill useful functions, yet there is limited transparency about the design details, such as how their behaviors, embodiments, and workflow mappings were determined. Prior studies also show that some deployments can misalign with daily practice when embodiment details (e.g., screens, voices, affordances) fail to fit the environment \cite{ren2024working}. By foregrounding how robot behaviors and appearances align with workflow logic, our co-design process anticipates these challenges earlier in the design process.

\paragraph{2) Capability-aware envisioning.}
Emerging embodied-AI capabilities, such as multimodal reasoning, empathetic dialogue, and mobile manipulation, offer new directions for care collaboration. While past systems (e.g., pediatric ED distraction robots) were often remotely operated and scripted with limited interactivity \cite{hudson2023perspectives}, current trajectories support more responsive and autonomous behaviors. Our participants envisioned robots that could welcome patients, answer questions, deliver and sanitize toolkits, and offer emotional support. These visions drew on expressive modalities (e.g., lighting, gesture), richer embodiment materials, and dialogic abilities that support responsive and socially aware engagement.

\paragraph{3) Cross-context learning across healthcare service types.}
Our comparison of ED, SDC, and LTR reveals how shared care goals, such as helping patients feel oriented, reassured, or heard, manifest differently depending on the service nature.
The ``tour guide'' role designed by our participants, for example, appears as directional hallway assistance in ED, environmental orientation in SDC, and is less emphasized in LTR where residents are familiar with the space. Similarly, our artifacts also indicate that delivery tasks differ by urgency and storage constraints. ED demands precise, time-sensitive supply drops while SDC emphasizes larger storage to support sleep study setup, with less urgency.
These nuanced distinctions show how even seemingly similar roles require facility-specific interpretation and adjustment, which is knowledge that only emerges through comparative, cross-context design.

\subsection{Toward Considerate Embodied AI}
Embodied AI is increasingly envisioned as a technological paradigm with the potential to enter everyday life and empower human activities across critical domains such as healthcare \cite{liu2025survey}, education \cite{memarian2024embodied}, industry \cite{xu2025embodied} and home environments \cite{tsui2025exploring}. 
Unlike virtual AI systems that operate primarily in digital environments (e.g., language-model-based chatbots that do not directly interact with the physical world), embodied AI enters human environments from the moment it is deployed. 
This compels us to ask: \textit{In a future where embodied AI systems move from isolated prototypes to widespread everyday presence, no longer rare tools in specialized settings but ubiquitous companions, tutors, and colleagues, how can we design them to be meaningfully integrated into our social worlds?}

We argue that this capacity for coexistence is not a peripheral concern, but a fundamental precondition for acceptance, trust, and long-term integration. 
To address this, we must design \textit{considerate} embodied AI: systems that go beyond functional adequacy to cultivate sensitivity to social norms (e.g., respecting turn-taking in conversations or privacy boundaries), spatial dynamics (e.g., navigating crowded hallways without disrupting human flows), and emotional needs (e.g., offering calming cues and empathetic tone in stressful care settings). 
This is especially critical in high-stakes domains such as healthcare, where misalignment with professional practice can result in friction, or even harm \cite{taylor2024towards}.

\subsubsection{Designing for Shared Routines and Relational Boundaries}
Participants explored how robots might move from being perceived as mere \textit{tools} to becoming \textit{relational actors} in everyday care \textit{routines and relationships}.
Long-term co-design prompted participants to reflect on spatial and interactional concerns rooted in daily routines and professional practice, during which participants raised new and evocative questions. 
For instance, in the LTR context, P5 remarked, \textit{``If I were a robot, I wouldn't want someone to grab my pants''}, highlighting how participants imbued robots with a kind of embodied sensitivity, imagining what it might mean for a robot to have its own spatial or relational boundaries. 
This resonates with prior work that similarly examines how relational expectations are placed on robots, exploring how cancer care robots may be expected to express compassion and bear witness to human distress, while simultaneously questioning the relational boundaries and responsibilities such expectations entail \cite{jayaraman2026designing}.

\subsubsection{Prototypes as Instruments of Thought}
To explore how robots might participate in everyday care routines, we found that lower-fidelity artifacts, though useful for early brainstorming, could not capture the intricacies involved. 
Therefore, \textit{high-fidelity prototyping} is necessary not only to refine ideas but also as a \textit{thinking apparatus} for revealing real-world deployment complexities, such as ergonomics, safety, workflow integration.
High-fidelity prototyping remains relatively uncommon in healthcare HRI, with prior work including examples that focus on the rapid prototyping of modular robots, such as sanitation or social interaction modules, adapted iteratively to align with user needs \cite{colle2025co}.
Our study resonates with this trajectory, but also details how prototypes across stages shaped participants' involvement and deepened their understanding of robots in practice. 
We argue that evolving into high-fidelity prototyping was valuable not only for refining solutions, but also for cultivating familiarity with robotics and thereby facilitating their acceptance in integrated care environments.

\subsubsection{Designing Across Disciplines and Sensibilities}
These insights were further shaped by \textit{multidisciplinary collaboration} among HCWs, caregivers, engineers, and artists. 
Each group brought unique perspectives to the table.
HCWs and caregivers foregrounded workflow and patient needs, engineers mapped feasibility constraints, while artists emphasized sensory and affective qualities (e.g., how illumination or material choices could create calming atmospheres in the SDC). 
For example, resonating with our findings, an editorial on arts and robotics noted that artistic engagement can broaden the expressive and social potential of robot design \cite{herath2022art}.
These heterogeneous perspectives collectively grounded designs not only in functional logic but also in experiential and emotional resonance, producing robots shaped by both practical feasibility and lived experience.

\subsubsection{Envisioning Collaborative Robot Ecologies}

We also noted a broader tendency that cut across teams and contexts: participants often envisioned multiple robots distributed across a context, each playing distinct yet coordinated roles such as escorting patients, delivering supplies, or gathering contextual data to support care planning. 
What emerged was not merely a set of individual assistants, but a paradigm of \textit{multi-robot collaboration} embedded in care processes, where robots are conceived as coordinated actors alongside human and non-human teammates.
Our findings are supported by recent research on this topic. 
For example, studies have investigated how people perceive robots in staff hierarchies \cite{abrams2025teaming}, how different communication modalities impact teamwork in time-sensitive healthcare contexts \cite{tanjim2025human}, while others have explored the technical challenges of coordinating robot teams \cite{bai2025virtual}.

\subsection{Limitations and Future Work}

Our study has several limitations. (1) Recruitment and Participation. Because the study required sustained commitment, only five HCWs from the targeted facilities were able to participate consistently (ED: 2; LTR: 1; SDC: 2), which may affect representativeness. In addition, we documented progress at the group level rather than tracking individual attendance week-by-week, which limits fine-grained analysis of individual participation patterns. Further, the high-commitment and uncompensated nature of the workshop may have attracted participants who were more intrinsically enthusiastic about robotics, potentially limiting attitudinal diversity.
(2) Robotic Platform Constraints. 
We used TIAGo primarily as a mobile base to anchor participants' thinking about
navigation and movement while allowing them to imagine new embodied forms on
top of it. However, relying on a single platform may have
constrained participants' exploration of other emerging embodied-AI
possibilities (e.g., humanoids, soft robotics, XR-integrated robots).
(3) Contextual Coverage.  
Time constraints prevented full coverage of each setting. ED discussions focused
on walk-in flows rather than ambulance arrivals; SDC sessions centered on
in-clinic studies rather than in-home testing; and LTR concepts reflected the
residents and art-therapy context. All three sites were in the Global North,
which may limit transferability to other care models.
(4) Group Dynamics and Social Factors.  
Although interviews reflected cooperation, in-session observations revealed
moments of tension, for example, disagreements within the SDC team about how the drawer should open and how its door should be designed for nighttime tasks. Furthermore, our longstanding
relationships with participants may have introduced social-desirability bias,
with more critical views less likely to be voiced.

Future work will expand this approach to additional facility types, geographic regions, and embodied-AI platforms, and will explore multi-robot coordination in more complex healthcare workflows. Future studies could also examine how participants' attitudes toward embodied AI evolve and shape design outcomes, as well as how role distribution and relative group size influence multidisciplinary collaboration. Finally, future work may incorporate more systematic attendance tracking across long-term co-design efforts.

\section{Conclusion}

Designing embodied AI for healthcare requires integrating diverse stakeholder needs and expertise, yet prior work rarely brings together multidisciplinary collaboration, iterative progression toward high-fidelity prototypes, and support for non-technical participants within a sustained co-design process.
We conducted a 14-week multidisciplinary co-design workshop with 22 participants to design healthcare embodied AI addressing NVA work across three distinct care settings (LTR, ED, and SDC), scaffolded by educational sessions covering technical and design foundations.
We analyzed multi-stage design artifacts (storyboards, cardboard prototypes, and full-scale prototypes) alongside interviews and post-educational session discussions.
Our findings highlighted the need to align designs with facility-specific priorities while remaining attentive to contextual nuance. 
Sustained multidisciplinary collaboration fostered mutual learning; iterative prototyping scaffolded shifts from abstract brainstorming to embodied thinking; and educational scaffolds helped non-technical participants make design decisions with awareness of real-world constraints while also broadening technical participants' perspectives.
Beyond functionality, our study highlights that the central challenge of embodied AI is meaningful coexistence in human environments, sharing routines, respecting relational boundaries, and becoming trusted members of sociotechnical teams.
To advance this vision, we distilled eight design guidelines for developing more \textit{considerate} embodied AI systems that are not only technically feasible, but also socially acceptable, trustworthy, and meaningfully integrated into the everyday ecologies.

\begin{acks}
This material was supported by the National Science Foundation under Grant No. IIS-2423127.
\end{acks}

\bibliographystyle{ACM-Reference-Format}
\bibliography{main.bib}

\appendix

\section{Workshop Activities Details}
\label{app:activities_detail}

\textbf{Pre-Workshop Session. } One week before the workshop series began, we hosted a pre-workshop introduction session to discuss the goals of the workshop series, participants' expected contributions to activities, and expected outcomes at the end of the workshop.
We started by introducing the research team.
Then, we discussed the goals to integrate advanced robots into ED, LTR, and SDC facilities to handle repetitive tasks and support HCWs to focus more on patient care.
Then, we summarized the anticipated workshop activities.
The expected workshop outputs are robot prototypes with unique forms for the 3 aforementioned healthcare facilities.

\textbf{Workshop Introduction. } Building on prior work \cite{ostrowski2021long}, we employed a co-design process that combined community-building activities, reflection on setting-specific challenges, and robot ideation, along with hands-on equipment training to prepare participants for later robot prototyping and educational content on HRI.
We started by administering a consent form and ``Pre-workshop Survey'' to collect demographic information of the participants and iteratively administered these forms to new participants who engaged in the workshop over three months.
Next, we engaged in an icebreaker activity called, ``What is a Robot To You?'', which involved participants drawing a picture of what they imagined a robot looks like.
Then, we engaged in a reflection discussion where participants shared their drawings and reflected on what they perceived a robot to be, how it behaves, and interacts with people.

\textbf{Stakeholder Panel Discussion, Training, and Robot Demonstration. } We began the workshop with a stakeholder panel, which guided us through \textit{empathy mapping} to uncover the problems and needs of our stakeholders and ideate on the potential for robots to help. 
One worker from the ED, SDC, and LTR facility sat on the panel.
The ED panelist is a Physician with 10-15 years of experience, LTR panelist is a certified therapeutic recreation specialist with more than 20 years of experience, and a certified dementia practitioner, and the SDC panelist is a Sleep Medicine Physician with more than 20 years of experience.
Some panel questions include: ``What is your role and walk us through an average work day?'',
``Are there certain tasks in your job that you believe do not require a skilled worker or keep you away from patients?'',
``How do these tasks pose challenges that might negatively impact your work?'',
``Who should the robot interact with?''.
Then, participants engaged in safety orientation and robot demonstration.
The safety training involved an introduction to MakeLab equipment (e.g., 3D printer, laser machine) and safety protocols.
Following the safety training session, we introduced the TIAGo Mobile Base robot platform in subsequent sessions to give participants hands-on exposure to the robot’s capabilities and real-world scale. 
This session concluded with introducing participants to the TIAGO robotic base platform and exploring possible physical extensions that could be developed to enhance its functionality. 
This provided participants with a clearer understanding of how the robot could interact within the space and how its design could be adapted to meet specific needs.
We described the sensing, perception, navigation, and interaction capabilities of the robot (see Section \ref{sec:robot}).
Using a teleoperation web-based application built by our team \cite{arias2025caris}, we demonstrated the robot by moving it around, used text-to-speech software to enable the robot to talk to participants, and used a Large Language Model to demonstrate question answering scenarios between the robot and participants (i.e., ``TIAGo, what do you see in the environment?'').

\textbf{Introduction to Laser Cutting and HRI. } The second workshop started with a recap of the running notes that summarize the first session, followed by a laser cutting activity and a healthcare site model activity.
As participants joined the workshop, they engaged in a laser cutting ``Name Tag'' exercise that involved etching on 3”x1” metal with their name and role (i.e., stakeholder, maker, programmer).
This activity allowed participants to create personalized, wearable name tags that they used throughout the workshop.
We trained participants using Adobe Illustrator software to do the vector diagram design and Epilog Laser Machine to do the laser cutting.
To familiarize participants with introductory knowledge of robots, we invited a professor at the university who specialized in HRI and medical robot design.
They presented fundamental concepts and background information about HRI, including the importance of robot design for effective integration in healthcare settings, and current use cases of robots in healthcare.

\textbf{Dollhouse Activity. } Based on insights from the stakeholder panel, it was crucial to capture individual panelists' reflections on the proposed problems for each healthcare site.
To support this, we designed a “Dollhouse Activity”. 
First, we gathered floor plans from each organization, which were then imported into 3D modeling software SketchUp for tracing over to create a 1:1 scale CAD model. 
It was important to gain the same scale grid over each plan to have comparable 3D printed pieces.
On the day of the activity, participants were divided into three groups, one for each healthcare facility type.
A stakeholder was assigned to their respective 3D facility model and presented with a printed floor plan layered with a transparent film and 2 color markers. 
One to showcase patient's path of movement vs. another color to showcase the healthcare giver’s path of movement in that same space.
Participants also had 3D printed furniture pieces, which they could use to create scenarios on the 2D plan, allowing actual users of the space to demonstrate their movement paths and suggest potential routes for robot navigation in real time to other participants.

\textbf{Co-creation through Storyboarding. } To further help visualize how robots could interact with stakeholders in real-world scenarios, we began storyboarding exercises. 
The goal of this activity was for participants to select a scenario for each space and create a storyboard that visualizes a scene where a robot interacts with humans. 
Participants were encouraged to focus on illustrating the types of interactions and the environment where these interactions occur.
Each group was assigned a whiteboard and tasked with drawing out potential robot usage by imagining various scenarios in 6-9 squares. 
After completing their storyboards, groups shared their concepts with one another, serving as their first design draft, reflection, and negotiation of important design features. 
This collaborative step allowed participants to refine their ideas and better understand interaction possibilities from multidisciplinary perspectives.

\textbf{Inclusive Paper Prototyping Activity. } To further bridge the gap between design and implementation, we provided laser-cut cardboard miniature versions of the TIAGO robot base to each group. 
After receiving training, participants used available materials and tools to quickly create small-scale versions of the robot embodiment that would sit atop the base, including paper, straws, glue, fabric, and more (see Figure \ref{fig:artifacts}). 
This hands-on building exercise will serve as a foundation for the next step in prototyping the robot’s embodiment and will guide discussions around scale, interactions, and movement within the space.
We encouraged workshop participants to reflect on the robot paper prototypes and small group discussions.
Additionally, we invited professional craftsmen to deliver lightning talks, sharing valuable insights and essential prototyping skills to support participants in building their prototypes effectively.
Building on the work \cite{moharana2019robots}, participants reflected on the robot’s role in interactions, capabilities, form factor, activities to engage in, and appropriate interaction modalities to promote effective communication between robots and people.

\textbf{Full-Scale Prototyping Activity. } Building on the prior cardboard paper prototyping phase and preparatory efforts, including laser cutting training, expert-led educational sessions, and technical consultations, participants proceeded to construct full-scale prototypes of their envisioned robots. 
Each team received support from researchers, maker experts, and domain specialists who provided guidance throughout the process. 
While there were no constraints on the number of prototypes, all three groups focused on building one high-fidelity prototype due to time limitations. 
Prototyping extended beyond scheduled workshop hours, with participants returning to the lab for additional build time ahead of the final public showcase. 
These prototypes embodied earlier design ideas while integrating detailed functional features, including movable drawers, adjustable structural components, and interaction mechanisms grounded in specific task roles. 
Participants articulated their envisioned use contexts and interaction logic through in-person discussions, follow-up emails, and informal reviews, providing rich justifications for their design decisions.

\textbf{Reflection Interviews. } The goal of the reflection interviews was to explore participants' thoughts on the stakeholder panel session, multidisciplinary collaboration experiences, and the workshop organization in general. 
We conducted one-on-one formal interviews after 4 weeks of the workshop, based on participants’ availability and willingness to ensure a respectful and comfortable process. 
Among the participants in the two workshops, we inquired about their availability for approximately 20-minute interviews during the prototyping sessions. 
Participants who were occupied with the prototyping were not pressured to participate. Additionally, we prioritized willingness to ensure a positive experience. 
For instance, one participant (P20) declined the interview but suggested another participant (P19) who might provide more insights. 
As a result, we conducted 11 interviews, inviting participants to a quiet corner to facilitate a focused discussion.
While we prepared prompting questions, we encouraged open-ended sharing beyond these prompts to capture richer insights. 
The prompting questions include ``Which parts of the workshop do you find interesting or rewarding and which parts need improvement?'', ``What suggestions do you have for increasing the flexibility of the workshop?'', and “What are your thoughts on the concept of role-switching, which means participants may have and contribute to different roles?”

\textbf{Educational Session - ``Let’s Talk More About Robotics''. }
This educational session introduced participants to foundational knowledge about robotics, particularly in healthcare contexts. 
We began by tracing the historical origins of the term ``robot” and showcasing landmark developments, such as the first humanoid designs and the emergence of robots capable of environmental reasoning. 
Next, we explored a range of healthcare robot applications, from surgical assistance and in/outpatient support (e.g., delivering blankets, sanitizing spaces) to rehabilitation in long-term care facilities. 
These examples grounded our discussion of the real-world usability, safety, accessibility, and implementation challenges surrounding healthcare robotics. We then introduced the basics of robot programming using ROS2 \cite{ros}. 
Finally, we emphasized the importance of considering human preferences in robotic systems design, using robotic arm trajectory as a concrete case to illustrate how individual preferences can vary.

\textbf{Educational Session - ``Controls and Operating Systems”. }
We introduced participants to robot control and operation. We began by demonstrating various ways to robot navigation: For example, it can be operated manually using a joystick or keyboard (e.g., assigning specific commands to interface buttons), or programmed to respond automatically to environmental stimuli such as humans or obstacles. We also introduced the concept of mapping an environment and setting target positions. These examples helped participants consider how robots perceive and navigate space. To prompt further reflection on robot autonomy and control, we facilitated a set of guiding questions to scaffold their design thinking: For example, ``How will your robot move?'', ``What are the surroundings like?'', ``Are any sensors needed?'', ``Will the robot rely heavily on sensors?'', ``What data does your robot need to perform as intended?''.
We concluded the session with a demonstration of a robot interface, featuring barcode scanning and language selection functionalities. To demonstrate potential use cases to participants, we implemented a series of scenarios to illustrate how human–robot interaction can be designed to meet different needs in healthcare settings. For example, for ED, users could verify their identity, check wait times, and request emergency assistance. For SDC, the interface provided educational videos, visit overviews, and Frequently Asked Questions (FAQs). For LTR, users could request personalized services such as meals or medical help, and access entertainment content. These examples helped participants consider how interface design can support more accessible, efficient, and context-aware interactions with robotic systems.

\textbf{Final Public Showcase. }
At the end of the 14-week workshop, we presented the co-ideation journey at a university-led public event, centered on the theme of how emerging technologies can shape societal impact, and attended by students, researchers, educators, entrepreneurs, and industry professionals.
We presented major milestones, design pivots, and collaborative moments through a visual timeline and storytelling. The full-scale prototypes were displayed alongside large-format posters detailing the design rationale, component functionalities, and intended use contexts.
To help visitors engage with the envisioned care scenarios and user interactions, we prepared an interactive interface simulating real-time robot control, accompanied by pre-recorded videos demonstrating how specific tasks were triggered through the interface.

\section{Artifacts Details}
\label{app:artifacts}
\subsection{Storyboard Results}

 Our workshop participants created and reflected on storyboards that depict how they envision robots interacting with workers, staff, and patients to engage in assistive activities in the SDC, ED, and LTR, as illustrated in Figure \ref{fig:artifacts}.

\textbf{Emergency Department. }For the ED scenario, participants based the story on one member’s experience as a patient who cut her finger. 
The HCW representative clarified that the scenario assumes a patient walking in, rather than arriving by ambulance. 
The storyboard follows the patient’s journey from registration through treatment and discharge. 
They propose six distinct types of robots to assist throughout the above healthcare process:

At the registration desk, the \textbf{Registration Robot} facilitates patient check-in by guiding them through the process and assisting with language selection. 
It ensures that all collected information is securely transmitted to HCWs, enabling them to analyze the data and prepare for treatment promptly. 
This streamlines administrative tasks and increases efficiencies during registration.

In the waiting area, the  \textbf{Waiting Room Robot} engages patients with its friendly appearance and ability to communicate in their chosen language. 
It verifies patient identities, provides updates on estimated wait times, reports lab test status, and monitors patient well-being. 
By alerting healthcare workers in emergencies, it adds a layer of safety and reassurance, creating a more informed waiting experience for patients. 

Navigating large healthcare facilities can be burdensome, and the \textbf{Guiding Robot} addresses this by leading patients between different areas. 
With its compact design, it moves efficiently through crowded hallways, minimizing congestion and ensuring timely navigation for patients.

To support healthcare workers, the \textbf{Supply Robot} delivers essential medical supplies to where they are needed. 
For instance, when a doctor requests a laceration kit, the robot ensures rapid and accurate delivery, reducing the workload on staff and maintaining operational efficiency.

At the end of the care process, the  \textbf{Discharge Robot} plays a crucial role in ensuring patients are prepared to leave the facility. 
It collects necessary discharge details (e.g., asking, \textit{“Can I take a picture of your injury for records?”}, provides follow-up care instructions  (e.g.,  \textit{“Your sutures will need to be removed in 10 days”}), and offers assistance (e.g., arranging a ride or providing refreshments if needed). 
It can thus make the discharge from the facility smoother and more supportive for patients.

Lastly, the \textbf{Comfort Robot} enhances the care environment in waiting and discharge areas by offering snacks and drinks to both HCWs and care recipients. 
It promotes a more pleasant and accommodating atmosphere for all.
These robotic solutions not only address specific functional needs but also help create a more supportive environment for both care recipients and HCWs.

\textbf{Long-term Rehabilitation Facility. } In the LTR scenario, participants developed a narrative centered around a typical day for a resident where the resident wants to  \textit{“socialize or get some coffee in the day room (a room in the facility)”}. 
They highlighted four main capabilities of the robot.
It begins by \textbf{Identity Detecting} when called by a resident and logs them in automatically, ensuring secure and smooth access to the system. 
Once logged in, the robot \textbf{Retrieves Schedule} for the resident, keeping them informed of upcoming activities.
Beyond schedule management, the robot also \textbf{Provides Entertainment}. Residents can request games, such as puzzles, helping to engage them in enjoyable ways. 
Additionally, the robot \textbf{Tracks and Records Interaction Data for Therapy}, which is shared with the art therapist at the end of the day. 
This data provides insights into the residents’ engagement levels and can inform personalized care strategies or therapeutic activities.
By automating this process, the robot relieves staff from manual record-keeping and enables more timely, informed decisions regarding care.
Throughout the day, the robot interacts with residents in a friendly and accommodating manner. 
If multiple residents want to request the same information, they can wait for their turn to interact with the robot. 
This robotic solution frees up HCWs’ time and helps improve residents’ overall well-being by keeping them engaged and socially connected within the facility.

\textbf{Sleep Disorder Clinic. } For SDC, participants envisioned a patient’s journey through the entire sleep therapy process and proposed nine distinct tasks where robots could assist in the healthcare workflow:
The journey begins with the \textbf{Concierge Robot}, which warmly greets patients and verifies their identification, ensuring a seamless check-in process. 
Following this, a \textbf{Tour Guide Robot} leads patients through the clinic’s common areas, offering a comfortable introduction to the space.
Once settled, a robot provides \textbf{General Orientation}, including basic information about the sleep therapy process, helping patients understand what to expect during their stay. 
It can also \textbf{Personalized Education}, delivering tailored educational content specific to each patient’s condition or specific needs, enhancing their understanding of sleep therapy.
During overnight stays, a \textbf{Midnight Questionnaire Robot} distributes questionnaires to gather patient feedback or collect required data. 
This ensures that essential information is collected without adding to the workload of healthcare workers during late hours.
Behind the scenes, a \textbf{Technician Support Robot} assists technicians by managing inventory and delivering necessary supplies to the control room, optimizing workflow for technicians. 
A \textbf{Task Reminder Robots} also sends reminders to technicians about scheduled tasks, helping them stay on track throughout the day.
Additionally, a \textbf{Data Management Robot} identifies and collects missing data from relevant sources, ensuring that patient records are complete and up-to-date. 
Finally, as patients complete their journey and prepare to leave, an \textbf{Exit Assistance Robot} guides them to the exit while gathering feedback on their overall experience, providing valuable insights for clinic improvement.
These robotic solutions streamline various stages of the sleep therapy process, reducing the workload for HCWs and enhancing efficiency compared to the manual handling of these tasks.
\subsubsection{Cardboard Prototype}

Building on storyboarding, participants created cardboard prototypes, iterated and refined their ideas, and freely collaborated within their groups.
To facilitate the prototyping process, we provided several attributes for consideration: robot name, role of the robot, features \& activities, and interaction modalities.
The LTR, ED, and SDC teams built 2, 3, and 2 prototypes, respectively. The details of the attributes are listed in Figure \ref{fig:artifacts}.

\subsubsection{Full-Scale Prototype}

Building on these mid-fidelity designs, each team then advanced to constructing one full-scale prototype per setting. The details of the attributes are listed in Figure \ref{fig:artifacts}.

\end{document}